\documentclass[12pt]{article}
\usepackage{amscd,amssymb,amsmath,latexsym,enumerate}
\usepackage[mathscr]{euscript}
\usepackage{epsfig}
\usepackage{textcomp}
\usepackage{amsmath}
\usepackage{amssymb}
\usepackage{verbatim}

\title{Orbital polarization and magnetization 
\\
for independent particles in disordered media}
\vspace{.5cm}

\author{Hermann Schulz-Baldes$^{1}$, Stefan Teufel$^2$\\
{\small $^1$ Department Mathematik, Universit\"at Erlangen-N\"urnberg, D-91058
Erlangen, Germany}
\\
{\small $^2$ Mathematisches Institut, Universit\"at T\"ubingen, D-72076 T\"ubingen, Germany}
 \vspace{.2cm}
}

\date{ }

\newtheorem{theo}{Theorem}

\newtheorem{proposi}{Proposition}
\newtheorem{lemma}{Lemma}
\newtheorem{coro}{Corollary}

\newcommand{\Aa}{{\mathcal A}}
\newcommand{\Bb}{{\mathcal B}}

\newcommand{\Ee}{{\mathcal E}}

\newcommand{\Ll}{{\mathcal L}}
\newcommand{\Hh}{{\mathcal H}}
\newcommand{\Oo}{{\mathcal O}}
\newcommand{\Tt}{{\mathcal T}}
\newcommand{\Nn}{{\mathcal N}}

\newcommand{\BB}{{\bf B}}
\newcommand{\EE}{{\bf E}}

\newcommand{\CM}{{\mathbb C}}

\newcommand{\NM}{{\mathbb N}}
\newcommand{\PM}{{\mathbb P}}
\newcommand{\RM}{{\mathbb R}}
\newcommand{\SM}{{\mathbb S}}
\newcommand{\TM}{{\mathbb T}}
\newcommand{\ZM}{{\mathbb Z}}

\newcommand{\TR}{{\rm Tr}}

\def\esssup{\mathop{\rm esssup}}

\newcommand{\one}{{\bf 1}}

\def\XXint#1#2#3{{\setbox0=\hbox{$#1{#2#3}{\int}$}
     \vcenter{\hbox{$#2#3$}}\kern-.5\wd0}}

\begin{document}

\maketitle

\begin{abstract} 
Formulas for the contribution of the conduction electrons to the polarization and magnetization are derived for disordered systems and within a one-particle framework. These results generalize known formulas for Bloch electrons and the presented proofs considerably simplify and strengthen prior justifications. The new formulas show that orbital polarization and magnetization are of geometric nature. This leads to quantization for a periodically driven Piezo effect as well as the derivative of the magnetization w.r.t. the chemical potential. It is also shown how the latter is connected to boundary currents in Chern insulators. The main technical tools in the proofs are an adaption of Nenciu's super-adiabatic theory to C$^*$-dynamical systems and Bellissard's Ito derivatives w.r.t. the magnetic field.
\end{abstract}




\section{Introduction}

Polarization and magnetization are two important properties of solid state systems. In classical electromagnetism they are often described phenomenologically, but within the framework of quantum mechanics they ought to be calculated from first principle. It turns out that both are composed of two main contributions, a relatively simple one resulting from the ions and spins respectively and a more delicate one stemming from the conduction electrons. This latter contribution is called {\it orbital} and this work is about how to calculate it, at least in principle, for disordered solids within a one-particle model for which the interactions between electrons (or quasiparticles) are neglected. 

\vspace{.2cm}

Let us first focus on the polarization and the Piezo effect as the most prominent example. 
Deformations of a crystal can lead to a polarization of the sample, {\it i.e.} to the accumulation of charges on the boundary of the sample. The first contribution to this effect is the polarization due to relative displacements of the ionic cores in a unit cell of the crystal and it is straightforward to compute. The second contribution comes from the conduction electrons.  This contribution was only understood in the last 20 years. The key observations are, first of all, that the polarization of the sample is a bulk effect resulting from a non-vanishing current induced by a change of the Hamiltonian in time and, second of all, that only changes in the polarization are well defined. Based on a linear response argument, King-Smith and Vanderbilt derived  a formula for the polarization in terms of instantaneous Bloch eigenfunctions of the underlying periodic system \cite{KSV}. A review of the more recent physics literature is given by Resta \cite{Res}. The King-Smith and Vanderbilt formula was rigorously derived and generalized to continuous and periodic systems in \cite{PST}. More precisely, it was shown that a certain generalization is correct in the adiabatic limit of slow deformations without power law corrections.

\vspace{.2cm}

In this work, the focus is on independent electrons in a disordered medium and in a tight-binding approximation. The class of space-homogeneous Hamiltonians is described in detail in Section~\ref{sec-Hamiltonian}. Motivated by the above picture, let us suppose that there is a continuous path $t\in[0,T]\mapsto H(t)$ of time-dependent Hamiltonians. Then states evolve according to the Liouville equation, see Section~\ref{sec-timeevolve}. The change $\Delta P$ in polarization during a change of the system between time $t=0$ and $t=T$ is then defined by integrating the induced current density carried in the time-evolved state $\rho(t)$ of the system:
\begin{equation}
\label{eq-polardef}
\Delta P \;=\; \int_0^T dt\;  \Tt_\BB ( \rho(t) \,J(t)) \,.
\end{equation}
Here $J(t)=\imath[H(t),X]$ is the current operator given by the commutator of $H(t)$ and the position operator and  $\Tt_\BB$ is  the trace per unit volume at fixed magnetic field $\BB$. All these objects obtain a clear mathematical meaning in Section~\ref{sec-setup}. Note that $\Delta P$ is still a density and needs to be integrated over a suitable surface to compute the accumulated surface charge. Let us also point out that, even if the initial condition $\rho(0)$ is given by a function of the Hamiltonian, it is not true that also $\rho(t)$ is a function of the Hamiltonian (in which case the integrand in \eqref{eq-polardef} would vanish by Proposition~\ref{prop-zerocurrent} below). It is now relatively easy (see Proposition~\ref{prop-polarid}) to derive that, if $\rho(0)=P_0(0)$ is the Fermi projection on all states below the Fermi level $\mu(0)$ lying in a gap of $H(0)$, then $\rho(t)$ is also a projection and
\begin{equation}
\label{eq-KV0}
\Delta P\;=\;
\int^T_0 dt\;
\Tt_\BB\bigl(
\rho(t)\,[\partial_t \rho(t), [X,\rho(t)]]\bigl)
\;.
\end{equation}
This formula is not very useful because it requires the knowledge of the time evolution of $\rho(t)$ which is, in general, not a function of $H(t)$. However, in the adiabatic limit of time slow changes,  the situation is simpler as proved in our first main result.

\vspace{.2cm}

\noindent {\bf Theorem} (detailed version in Theorem~\ref{theo-polar} below)
{\it Let $t\mapsto H(t)$ be smooth with a gap that remains open in $t$ and let   $\frac{d}{d t}H(t)$ be compactly supported in $(0,T)$. The spectral projection below this gap is denoted by $P_0(t)$. For the   time-evolution governed by $\varepsilon\,\partial_t\rho(t)=\imath[\rho(t),H(t)]$ with initial condition $\rho(0)=P_0(0)$ one has for any $N\in\NM$
\begin{equation}
\label{eq-KV00}
\Delta P\;=\;
\int^T_0 dt\;
\Tt_\BB\bigl(
P_0(t)\,[\partial_t P_0(t), [X,P_0(t)]]\bigl)
\;\;+\;\;\Oo(\varepsilon^N)
\;.
\end{equation}
}

The leading order expression in \eqref{eq-KV00} is the above mentioned King-Smith and Vanderbilt formula. As already pointed out in the abstract, apart from \eqref{eq-KV0} the key ingredients to the proof is an adaption of  Nenciu's superadiabatic expansions to the C$^*$-algebra of homogeneous observables.  In the Corollaries~\ref{coro-diff} and \ref{coro-polar} it is shown that the r.h.s. of (\ref{eq-KV00})  is of topological nature. It does not change under diffeotopic deformations of the path $t\mapsto P_0(t)$ and, for a periodically  driven system of period $T$, it is qunatized, as noted by Thouless \cite{Tho} in a more restricted context.

\vspace{.2cm}

Next let us come to the magnetization of a solid. First of all, there is a contribution due to alignment of nuclear spins. This phenomena can be studied by a Heisenberg model. It may lead to a local magnetic field, which in turn can generate a second so-called orbital contribution resulting from the motion of the conduction electrons by the Biot-Savart law. If the electron Hamiltonian is time-reversal invariant, then the orbital magnetization vanishes (see Section~\ref{sec-magn}). If, however, there is a spin magnetization, the time-reversal invariance of the electron system is broken possibly resulting in orbital magnetization. Moreover, it may then be of comparable size, even though a secondary effect. Nevertheless, only quite recently there has been progress on how to calculate the orbital magnetization in the physics community. One reason for this may be that the magnetization cannot be calculated by linear response theory. The first (independent) works seem to be \cite{CTVR,XYFN} and they exhibited the orbital magnetization as a geometric quantity which can be expressed in terms of the Berry curvature and the Rammal-Wilkinson tensor of the Bloch bands. Since then there were numerous contributions which are well-documented in three recent reviews \cite{Res,XCN,Tho}.

\vspace{.2cm}

Now let us define the orbital magnetization of a system of independent electrons described again by  a (covariant family of) one-particle Hamiltonian $H$. One starts with the pressure (free energy per unit volume) at inverse temperature $\beta$ and chemical potential $\mu$:
\begin{equation}
\label{eq-pressuredef}
p({\beta,\mu})
\;=\;
\frac{1}{\beta}\;
\Tt_\BB\bigl(
\ln(\one+e^{-\beta(H-\mu)})
\bigr)
\;.
\end{equation}
Then the components of the magnetization $M({\beta,\mu})=(M_1({\beta,\mu}),\ldots,M_d({\beta,\mu}))$ are its derivatives w.r.t.\ the magnetic field:
\begin{equation}
\label{eq-magdef}
M_j({\beta,\mu})
\;=\;
\partial_{B_j}\,
p({\beta,\mu})
\;.
\end{equation}
Of course, one has to verify that the pressure is indeed differentiable. This is obvious within the formalism presented below, but can seem rather puzzling if one has in mind that a constant magnetic field $\BB$ asks for an unbounded vector potential for which there is little hope to control perturbation theory.  The same technical problem appears when trying to calculate the magnetic susceptibility, as explained in \cite{BCS}.  One way out is to use periodic magnetic fields and then study the long wave length limit \cite{ML,SVXN}. Here we follow a different path and view the magnetic field as parameter encoding the algebraic structure, more precisely, the type of non-commutativity of the magnetic translations. Therefore, the Hamiltonian is a single element in a field of algebras indexed by the magnetic field and within this field derivatives w.r.t.\ the magnetic field are readily defined. This idea was used by Bellissard \cite{Bel} to prove Streda's formula \cite{Str} as well as Lipshitz continuity of the boundaries of the Hofstadter butterfly \cite{Bel3}. A new key technical supplement to this formalism is a generalized DuHamel formula for the derivation w.r.t.\ the magnetic field (see Section~\ref{sec-derivativemag}).

\vspace{.2cm}

The first main result on the orbital magnetization (Theorem~\ref{theo-magT}) is that $M_j({\beta,\mu})$ can be expressed in a simple manner and without any approximations in terms of the current-current correlation measure (which also appears in the Kubo formula for the electrical conductivity \cite{KP,SBB}). From this we deduce the following new formula for the orbital magnetization.

\vspace{.2cm}

\noindent {\bf Theorem} (detailed version in Theorem~\ref{theo-magT=0} below) {\it Let the Fermi level $\mu$ lie in a spectral region of dynamical {\rm (}Anderson{\rm )} localization and let $P_\mu$ denote the Fermi projection. If the indices $j$ are understood cyclically in space dimension $d=3$, then the zero-temperature orbital magnetization is given by
\begin{equation}
\label{eq-magintro}
{M}_j({\infty,\mu})
\;=\;
\frac{\imath}{2}\;\Tt_{\BB}\bigl(|\mu-H|\,\bigl[[X_{j+1},P_\mu],[X_{j+2},P_\mu]\bigr]\bigr)
\;.
\end{equation}
}

For Bloch electrons and $\mu$ in a gap, it is shown in Section~\ref{sec-mag0} that this formula reduces to that of \cite{CTVR}. However, our compact formula \eqref{eq-magintro} is new and we believe the stated hypothesis to be necessary. Let us also stress that the presented derivation of \eqref{eq-magintro} is considerably more translucent than prior ones (apart from being rigorous and more general).  The expression on the r.h.s. of \eqref{eq-magintro} is somewhat similar to the formula for the Chern number in the quantum Hall effect \cite{BES}. Actually, the derivative $\partial_\mu {M}_j({\infty,\mu})$ is up to a numerical factor precisely equal to the Chern number as long as the Fermi level is in a region of dynamical localization (Theorem~\ref{theo-magderiv}). In particular, this derivative is topologically quantized to an integer and thus constant in any given interval of dynamical localization.
This shows that the magnetization is indeed of geometric nature.  

\vspace{.2cm}

In \cite{CTVR} it is already pointed out that boundary currents contribute to the orbital magnetization. Indeed, let us consider a two-dimensional sample in the form of a disc of radius $r$ which has a boundary current $I$ flowing around it. Then the associated magnetic moment is of order $I r^2$, so that the magnetization (magnetic moment per unit volume) is proportional to $I$. In Theorem~\ref{theo-boundary} it is stated that such boundary currents calculated from first principle in a system on a half-space are topologically quantized with the Chern number given by the bulk system. Therefore the strength of boundary currents is dictated by (the $\mu$-derivative of) the  magnetization itself or inversely the magnetization can be read off the boundary currents. This is in the spirit of topological insulators and similar to the situation in the quantum Hall effect \cite{KRS}.

\vspace{.3cm}

\noindent {\bf Acknoledgements:} H.~S.-B. thanks Baptiste Savoie for bringing the work of Resta \cite{Res} to his attention. S.\ T.\ thanks Gianluca Panati, Giuseppe DeNittis and Max Lein for numerous discussions on polarization and magnetization in the past. This work was supported by the DFG.

\section{Formalism}
\label{sec-setup}

In order to make this work essentially self-contained, we review the C$^*$-algebraic description of independent particles in homogeneous media \cite{Bel,Bel2}. Even though a bit lengthy, this is indispensable in order to fix notations. As already explained above, it is also a convenient framework for the study of variations in the magnetic field \cite{Bel2,Bel3}. 

\subsection{The Hamiltonian}
\label{sec-Hamiltonian}

The aim is to model electrons or holes (quasiparticles) in a solid which can be either periodic (so that Bloch-Floquet theory applies), quasiperiodic or random (as in the Anderson model). Let us suppose that the particles have a spin $s$ and are submitted to a periodic, quasiperiodic or random potential as well as a uniform magnetic field. There may also be spin-orbit interactions. The centers of the elementary cells of the underlying crystal will be identified with $\ZM^d$. We restrict ourselves to the tight-binding representation describing the low energy states. Hence the one-particle Hilbert space describing the quantum states is $\Hh=\ell^2(\ZM^d)\otimes \CM^L$ where $L=(2s+1)r$ and $r$ is the dimension of the other internal degrees of freedom ({\sl e.g.}, $r=2$ for the honeycomb lattice in $d=2$). The main focus will be on the cases $d=2$ and $d=3$. The constant external magnetic field is given by an antisymmetric real matrix $\BB=(B_{i,j})_{i,j=1\ldots d}$, namely $B_{i,j}= -B_{j,i}$. The magnetic translations
\begin{equation}
\label{eq-magntrans}
(U_a\psi)_n\;=\;
 e^{ \frac{\imath}{2} a\cdot \BB n }
\; \psi_{ n-a}
\mbox{ , }
\end{equation}
\noindent define a projective unitary representation of the {\em translation group} $\ZM^d$ on $\Hh$. Here the notation $a\cdot\BB n=\sum_{i,j}a_iB_{i,j}n_j$ is used. For dimension $d=2$ and $d=3$,  the magnetic field is parametrized as follows
\begin{equation}
\label{eq-parametrizeB}
\BB\;=\;
\begin{pmatrix}
0 & B_3
\\
-B_3 & 0
\end{pmatrix}
\;,
\qquad
\BB\;=\;
\begin{pmatrix}
0 & B_3 & -B_2
\\
-B_3 & 0 & B_1 
\\
B_2 & -B_1 & 0
\end{pmatrix}
\;.
\end{equation}
In dimension $d=2$, the magnetic flux trough a unit cell is $B_{3}$, and, for $d=3$, it is $B_{1}$, $B_{2}$ and $B_{3}$ in the directions $1$, $2$ and $3$ respectively. The exponential factors in \eqref{eq-magntrans} are explicitly given by
\begin{equation}
\label{eq-Bfactors}
e^{ \frac{\imath}{2} a\cdot\BB n }\;=\;
e^{\frac{\imath}{2} B_3(a_1n_2-a_2n_1)}
\;,
\qquad
e^{ \frac{\imath}{2} a\cdot\BB n }\;=\;
e^{\frac{\imath}{2} B_1(a_2n_3-a_3n_2)}e^{\frac{\imath}{2} B_2(a_3n_1-a_1n_3)}e^{\frac{\imath}{2} B_3(a_1n_2-a_2n_1)}
\;.
\end{equation}
As the magnetic field $\BB$ only enters all constructions below through the magnetic translations, it is parametrized by values in the torus $\Xi=[-\pi,\pi)^{\frac{1}{2}d(d-1)}$.

\vspace{.2cm}

The class of Hamiltonians of interest are strongly continuous families $(H_{\omega})_{\omega\in\Omega}$ of self-adjoint bounded operators on $\Hh$ indexed by a compact space
$\Omega$ of disorder or crystaline configurations. This space is supposed to be compact and furnished with a continuous action $T$ of the translation group $\ZM^d$. The family of Hamiltonians is then supposed to satisfy the covariance relation
$$
U_a
 H_{\omega}
  U_a^{-1} =
   H_{T^a \omega}
\mbox{ , }
\qquad a\in\ZM^d
\mbox{ . }
$$
In a typical situation, $H_\omega=H_0+V_\omega$ is the sum of a translation invariant kinetic operator $H_0$ and a potential energy $V_\omega$ of the form
\begin{equation}
\label{eq-Andmod}
H_\omega\;=\;H_0+V_\omega
\;,
\qquad
H_0\;=\;
\sum_{m\in\ZM^d} T_m\,\widetilde{U}_m\;,
\qquad
V_\omega\;=\;\sum_{n\in\ZM^d} V_{\omega,n}\;|n\rangle\langle n|
\;.
\end{equation}
Here the sum over $m$ has only finitely many terms and the coefficient matrices $T_m\in\mbox{\rm Mat}(L,\CM)$ satisfy $(T_m)^*=T_{-m}$ and $(\widetilde{U}_m\psi)_n= e^{ - \frac{\imath}{2} m\cdot\BB n }  \psi_{n-m}$. Furthermore, the matrices $V_{\omega,n}\in\mbox{\rm Mat}(L,\CM)$ are self-adjoint and continuous in $\omega$, and we used the bra-ket notations to denote by $|n\rangle$ the partial isometry in $\Hh$ onto $L$-dimensional state space supported by $n\in\ZM^d$. Then, if the $V_{\omega,n}$ are drawn from a compact set $K\subset\mbox{\rm Mat}(L,\CM)$, the set $\Omega$ is simply the Tychonov space $K^{\times\ZM^d}$ furnished with the shift action. It is also possible to construct $\Omega$ as the {\em hull} of a single given Hamiltonian $\widetilde{H}$ on $\Hh$ \cite{Bel}, namely as the strong closure of $\{ U_a\widetilde{H} U_a^{-1}\,|\, a\in \ZM^d \}$.  Then the Hamiltonian $\widetilde{H}$ is called {\em homogeneous} whenever the hull is compact. 

\vspace{.2cm} 

\subsection{The C$^*$-algebra $\Aa$ of observables}

Now let us construct an abstract observable C$^*$-algebra $\Aa$ the representations of which contain all the operators $H_\omega$, but also functions of $H_\omega$ as well as commutators of $H_\omega$ with the position operators and many more interesting operators which are space-homogeneous. All magnetic fields $\BB$ varying in the torus $\Xi$ appear as a parameter in this algebra which allows to study variations in $\BB$. The C$^*$-algebra $\Aa=(\Aa_\BB)_{\BB\in\Xi}$ is continuously fibered by the magnetic field, with each fiber $\Aa_\BB$ simply being a (reduced) twisted crossed-product of the action of $\ZM^d$ on $\Omega$. One advantage of this approach is that physically interesting objects can compactly be written out with derivatives and integrals that are automatically well-defined. In the case of periodic solid, all formulas reduce to standard textbook expressions of Bloch theory.

\vspace{.2cm}

The construction of $\Aa_\BB$ goes as follows. One endows the topological vector space $C_c(\Omega\times\ZM^{d}, \mbox{\rm Mat}(L,\CM))$ of continuous functions with compact support on $\Omega \times \ZM^{d}$ and values in $\mbox{\rm Mat}(L,\CM)$ with a $*$-algebra structure:
\begin{equation}
\label{eq-staralg0}
AB ({\omega}, n) \;=\;
 \sum_{l\in\ZM^{d}}\,
   A(\omega, l) \,B(T^{-l} \omega, n-l)\,
    e^{ \frac{\imath}{2} n \cdot\BB l}
\mbox{ , }
\qquad
     A^*(\omega, n) \; = \;
      A(T^{-n}\omega, -n)^*
\mbox{ , }
\end{equation}
\noindent where the last $*$ denotes the adjoint matrix. For $\omega \in \Omega$, a representation of this  $*$-algebra on $\Hh$ is given by
\begin{equation}
\label{eq-repreg}
 \bigl(\pi_{\BB,\omega}(A)\psi\bigr)_n\;=\;
  \sum_{l\in\ZM^{d}}
    \,A(T^{-n} \omega, l - n)\;
     e^{ \frac{\imath}{2} l\cdot\BB n } \;
       \psi_l
\mbox{ , }
\hspace{1cm}
       \psi \in \ell^{2}(\ZM^{d})\otimes\CM^L
\mbox{ . }
\end{equation}

\noindent The representations $\pi_{\BB,\omega}$ are related by the covariance
relation
\begin{equation}
\label{eq-covariance}
U_a\pi_{\BB,\omega}(A)U_a^{-1} \;=\;
 \pi_{\BB,T^{{a}}\omega}(A)
  \mbox{ , }
   \qquad a\in\ZM^d
\;,
\end{equation}
and are strongly continuous in $\BB$ and $\omega$. Now $\| A\| = \sup_{\omega \in \Omega}\| \pi_{\BB,\omega}(A)\|$ defines a $C^*$-norm on  $C_c(\Omega \times \ZM^{d},\mbox{\rm Mat}(L,\CM))$ and the observable C$^*$-algebra $\Aa_\BB=C(\Omega) \rtimes \ZM^{d}\otimes\mbox{\rm Mat}(L,\CM)$ is defined as the completion of $C_c(\Omega \times \ZM^{d},\mbox{\rm Mat}(L,\CM))$  under this norm.

\vspace{.2cm}

Now let $\Aa$ be the continuous field $(\Aa_\BB)_{\BB\in\Xi}$ of C$^*$-algebras in the sense of Dixmier \cite{Dix}. The compactly supported continuous matrix-valued functions on $\Xi\times\Omega\times\ZM^d$ are dense in the C$^*$-algebra $\Aa$ and their product and adjoint are then given fiberwise as in \eqref{eq-staralg0}:
\begin{equation}
\label{eq-staralg}
AB (\BB,{\omega}, n) \,=\,
 \sum_{l\in\ZM^{d}}\,
   A(\BB,\omega, l) \,B(\BB,T^{-l} \omega, n-l)\,
    e^{ \frac{\imath}{2} n \cdot\BB l}
\mbox{ , }
\qquad
     A^*(\BB,\omega, n) \, = \,
      A(\BB,T^{-n}\omega, -n)^*
\,.
\end{equation}
There exists a $*$-homomorphism $\rho_\BB:\Aa\to\Aa_{\BB}$ densely defined by evaluation at $\BB$ and we use the notation $\pi_{\BB,\omega}$ for $\pi_{\BB,\omega}\circ\rho_{\BB}$ . 

\vspace{.2cm}

Finally, let us exhibit explicitly an element $H\in\Aa$ the representations $\pi_{\BB,\omega}(H)$ of which are equal to the operator $H_\omega$ given in \eqref{eq-Andmod}. It is given by  the following continuous function on $\Xi\times\Omega\times\ZM^d$:
\begin{equation}
 \label{eq-Hexplicit}
H(\BB,\omega,n)
\;=\;
T_n\,+\,\delta_{n,0}\,V_{\omega,0}\;.
\end{equation}
%

\subsection{The derivations on $\Aa$}

On the C$^*$-algebra $\Aa$ exists a $d$-parameter group $k\in \TM^d\mapsto\rho_k$ of $*$-automorphisms defined by
$$
(\rho_k A)(\BB,\omega,n)
\;=\;
e^{\imath k\cdot n}\,A(\BB,\omega,n)
\;,
$$
where $k\cdot n=\sum_{j=1}^d k_jn_j$, namely $\rho_k$ is linear and satisfies $\rho_k(AB)=\rho_k(A)\rho_k(B)$, $\rho_k(A^*)=\rho_k(A)^*$ and the group property $\rho_k\rho_{k'}=\rho_{k+k'}$. As every $*$-automorphism, $\rho_k$ also conserves the C$^*$-norm. Its generators ${\nabla}=(\nabla_1,\ldots,\nabla_d)$ are unbounded, but closed $*$-derivations with domain $C^1_{ \rm s} (\Aa)$ which satisfy the Leibniz rule
$$
\nabla (AB)\;=\; (\nabla A)B+A(\nabla B)
\;,
\qquad
A,B\in C^1_{ \rm s}(\Aa)
\;.
$$
The index s indicates that only existence of space-wise derivatives is requested in $C^1_{ \rm s} (\Aa)$ (derivatives w.r.t. the magnetic field are dealt with further down). Explicitly the generators are given by 
\begin{equation}
\label{eq-nablader}
 \nabla_j A (\BB,\omega, n) \;=\;
  \imath\, n_j\, A (\BB,\omega, n)
\mbox{ , }
\qquad
A\in C^1_{ \rm s} (\Aa)\;.
\end{equation}
\noindent Note that these derivations are also defined on each fiber $\Aa_\BB$. Furthermore,  the sub-algebra $C^1_{ \rm s} (\Aa)\subset\Aa$ is endowed with the graph norm $\| A\|_{C^1_{ \rm s}} = \|A\|+ \sum_j \|\nabla_j A\|$ which makes it a Banach $*$-algebra as can readily deduced from the Leibniz rule. If the position operator ${X}=(X_1,\ldots,X_d)$ on $\Hh$ is defined as usual by 
$$
(X_j \psi)_n \;= \;n_j \psi_n\;,
\qquad
\psi=(\psi_n)_{n\in\ZM^d}\in\Hh\;,
$$
then 
\begin{equation}
\label{eq-derivrep}
\pi_{\BB,\omega}(\nabla_jA)
\;=\;
\imath [\pi_{\BB,\omega}(A),X_j]
\;.
\end{equation}
The following result is proved in \cite[Theorem 3.3.7]{Sak}, but for a slightly more regular function, it also follows from the proof of Proposition~\ref{prop-diffcalculus2} below.

\begin{proposi}
\label{prop-diffcalculus}
Let $H = H^*$ be an element of $C^1_{ \rm s} (\Aa)$. Then for any $f\in C^2 (\RM)$,  the operator $f(H)$ belongs to $C^1_{ \rm s} (\Aa)$. 
%
\end{proposi}

\vspace{.2cm}

\subsection{The trace per unit volume on $\Aa$}

Given a $T$-invariant probability measure $\PM$ on $\Omega$, a positive trace $\Tt_\BB$ on $\Aa$ (and each $\Aa_\BB$) is defined by
\begin{equation}
\label{eq-tracepervol}
 \Tt_\BB(A) \; = \;
  \int_{\Omega}
    \PM(d\omega)\;
    \TR\bigl(A(\BB,\omega , 0)\bigr)
\mbox{ . }
\end{equation}
The following is readily verified.

\begin{lemma}
\label{lem-trace} $\Tt_\BB$ is a linear functional defined on all of $\Aa$ and satisfies

\vspace{.1cm}

\noindent {\rm (i)} {\rm (}normalization{\rm )} $\Tt_\BB({\bf 1}) =  L$

\vspace{.1cm}

\noindent {\rm (ii)} {\rm (}positivity{\rm )} $\Tt_\BB(A^*A)\geq 0$ and $\Tt_\BB(A^*)=\overline{\Tt_\BB(A)}$

\vspace{.1cm}

\noindent {\rm (iii)} {\rm (}cyclicity{\rm )} $\Tt_\BB(AB)=\Tt_\BB(BA)$

\vspace{.1cm}

\noindent {\rm (iv)} {\rm (}norm bound\,{\rm )} $\Tt_\BB(|A B|)\leq \|A\|\,\Tt_\BB(|B|)$ where $|B|=(B^*B)^{\frac{1}{2}}$

\vspace{.1cm}

\noindent {\rm (v)} {\rm (}invariance{\rm )} $\Tt_\BB(\nabla A)=0$ for $A\in C^1_{ \rm s}(\Aa_\BB)$

\vspace{.1cm}

\noindent {\rm (vi)} {\rm (}partial integration{\rm )} $\Tt_\BB(A\nabla B)=-\Tt_\BB(\nabla A \,B)$ for $A,B\in C^1_{ \rm s}(\Aa_\BB)$

\end{lemma}

If $\PM$ is in addition ergodic, then Birkhoff's ergodic theorem implies that for any increasing sequence $(\Lambda_l)_{l\in\NM}$ of cubes centered at the origin
\begin{equation}
\label{eq-tracepervol2}
 \Tt_\BB(A) \; = \;
  \int_{\Omega}
   \PM(d\omega)\;
    \TR\bigl(\langle 0|\pi_{\BB,\omega}(A)|0\rangle\bigr)
     \;   =\;
      \lim_{l\rightarrow\infty}
       \frac{1}{|\Lambda_l| }
        \sum_{n\in\Lambda_l}
         \TR\bigl(\langle n| \pi_{\BB,\omega}(A) |n\rangle\bigr)
\mbox{ , }
\end{equation}

\noindent for $\PM$-almost all $\omega\in\Omega$. This shows that $\Tt_\BB$ is the trace
per unit volume.

\vspace{.2cm}

\subsection{The derivative w.r.t. the magnetic field}
\label{sec-derivativemag}

For sake of concreteness, we restrict to the cases $d=2$ and $d=3$ so that $\BB$ is parametrized by $B_j$, $j=1,2,3$, as given in \eqref{eq-parametrizeB}. It is straightforward to treat higher dimensions, but as it leads to heavier notations we refrain from doing so here. On 
$C^1_c(\Xi\times\Omega\times\ZM^d,\mbox{\rm Mat}(L,\CM))$, the set of all continuously differentiable matrix-valued functions with compact support, let us set for each $j=1,2,3$,
$$
(\delta_j A)(\BB)
\;=\;
\partial_{B_j}\,A(\BB)
\;.
$$
We also write $\delta=(\delta_1,\ldots,\delta_d)$. The operator $\delta_j$ is called {\it Ito derivative} in \cite{Bel2,Bel3}. It is not a derivation, but rather satisfies the following algebraic identities.

\begin{lemma}
\label{lem-deltajs} Let $d=2$ or $d=3$. Whenever all objects below are well-defined, one has:

\vspace{.1cm}

\noindent {\rm (i)} With the index $j$ taken $\mbox{\rm mod}\; 3$,
$$
\delta_j(AB)
\;=\;
(\delta_jA)B\;+\;A(\delta_jB)\;+\;
\frac{\imath}{2} \bigl(\nabla_{j+1}A\,\nabla_{j+2}B\,-\,\nabla_{j+2}A\,\nabla_{j+1}B\bigr)
\;,
$$

and
$$
\delta_j\,[A,B]
\;=\;
[\delta_jA,B]\;+\;[A,\delta_jB]\;+\;
\frac{\imath}{2} \bigl([\nabla_{j+1}A,\nabla_{j+2}B]\,-\,[\nabla_{j+2}A,\nabla_{j+1}B]\bigr)
\;.
$$

\vspace{.1cm}

\noindent {\rm (ii)} For an invertible $A\in\Aa$,
$$
\delta_j (A^{-1})
\;=\;
A^{-1}
\left(
-\delta_j A
+
\frac{\imath}{2}\, \nabla_{j+1}A\, A^{-1}\nabla_{j+2}A-\frac{\imath}{2}\, \nabla_{j+2}A\, A^{-1}\nabla_{j+1}A
\right)
A^{-1}
\;.
$$

\vspace{.1cm}

\noindent {\rm (iii)} $\delta_j(A^*)=\delta_j(A)^*$

\vspace{.1cm}

\noindent {\rm (iv)} $\delta_j\nabla_k A=\nabla_k\delta_j A$ 

\vspace{.1cm}

\noindent {\rm (v)} {\rm (}generalized DuHamel formula{\rm )} For $z\in\CM$ and index $j$ taken $\mbox{\rm mod}\; 3$,
\begin{eqnarray*}
\delta_j\, e^{zA}
& = &
z\int^1_0ds\,e^{(1-s)zA}\delta_j A\,e^{szA}
\\
& & \;\;\;+\;
\frac{\imath}{2} \,z^2\int^1_0ds\int^s_0 dr\;
e^{(1-s)zA}
\left(
\nabla_{j+1} A\,e^{rzA}\,\nabla_{j+2}A\,e^{(s-r)zA}
-\nabla_{j+2} A\,e^{rzA}\,\nabla_{j+1}A\,e^{(s-r)zA}
\right)
\end{eqnarray*}

\vspace{.1cm}

\noindent {\rm (vi)} $\partial_{B_j}\Tt_\BB(A)=\Tt_\BB(\delta_j(A))$

\end{lemma}

\noindent {\bf Proof.} (i) Let us focus on $j=3$. 
Deriving \eqref{eq-staralg} using \eqref{eq-Bfactors} shows
$$
\bigl(\delta_3 (AB)-(\delta_3 A)B-A(\delta_3B)\bigr)
 (\BB,{\omega}, n) \;=\;
 \sum_{l\in\ZM^{d}}\,
   A(\BB,\omega, l) \,B(\BB,T^{-l} \omega, n-l)\,
\frac{\imath}{2}\,(n_1l_2-n_2l_1)\,    e^{ \frac{\imath}{2} n \cdot\BB l}
\;.
$$
Thus replacing $n_1l_2-n_2l_1=(n_1-l_1)l_2-(n_2-l_2)l_1$ and then the definition of $\nabla_1$ and $\nabla_2$ allows to complete the calculation leading to the first identity in (i), which implies the second one. Then (ii) follows from $\delta_j (A^{-1}A)=0$. The points (iii), (iv) and also (vi) follow immediately from the definitions. The proof of (v) mimics that of DuHamel's formula, {\it e.g.} \cite[p.~68]{Sak}. For $n\geq 1$,
\begin{eqnarray}
\delta_j\,e^{zA}
\! \!\! & = &\!\!
\delta_j\,\bigl(e^{\frac{z}{n}A}\bigr)^n
\nonumber
\\
& = &\!\!
\delta_j\,e^{\frac{z}{n}A}\bigl(e^{\frac{z}{n}A}\bigr)^{n-1}
+
e^{\frac{z}{n}A}\,\delta_j\bigl(e^{\frac{z}{n}A}\bigr)^{n-1}
+\frac{\imath}{2}\, \nabla_{j+1}\, e^{\frac{z}{n}A}\,\nabla_{j+2}\bigl(e^{\frac{z}{n}A}\bigr)^{n-1}
-\frac{\imath}{2}\,\nabla_{j+2}\,e^{\frac{z}{n}A}\,\nabla_{j+1}\bigl(e^{\frac{z}{n}A}\bigr)^{n-1}
\nonumber
\\
& = &\!\!
\sum_{k=0}^{n-1}
e^{\frac{n-k-1}{n}zA}
\left(
\delta_j\,e^{\frac{z}{n}A}\;e^{\frac{k}{n}zA}
+\frac{\imath}{2} \,\nabla_{j+1}\,e^{\frac{z}{n}A}\;\nabla_{j+2}\,e^{\frac{k}{n}zA}
-\frac{\imath}{2} \,\nabla_{j+2}\,e^{\frac{z}{n}A}\;\nabla_{j+1}\,e^{\frac{k}{n}zA}
\right)
\;,
\label{eq-discretesum}
\end{eqnarray} 
where the third identity follows from iterating the second one. Now, with norm convergence,
$$
\lim_{n\to\infty}\, n\,\delta_j\,e^{\frac{z}{n}A}\;=\;z\,\delta_j A\;,
\qquad
\lim_{n\to\infty}\, n\,\nabla_j\,e^{\frac{z}{n}A}\;=\;z\,\nabla_j A
\;.
$$
Therefore the norm convergent Riemann integrals in \eqref{eq-discretesum} lead to
$$
\delta_j\,e^{zA}
\;=\;
\int^1_0 ds\;
e^{(1-s)zA}
\left(
z\,\delta_j A\;e^{szA}
+\frac{\imath}{2} \,z\,\nabla_{j+1}A\;\nabla_{j+2}\,e^{szA}
-\frac{\imath}{2} \,z\,\nabla_{j+2}A\;\nabla_{j+1}\,e^{szA}
\right)
\;.
$$
Finally replacing the remaining derivations of exponentials by their DuHamel formula and changing variables readily allows to complete the proof.
\hfill $\Box$

\vspace{.2cm}

As an illustration, let us point out that for $H=(H_\omega)_{\omega\in\Omega}$ in the form \eqref{eq-Andmod}, one has $\delta_j H=0$ because \eqref{eq-Hexplicit} is independent of $\BB$. As the products \eqref{eq-staralg} explicitly depend on $\BB$, this does not imply $\delta_j H^2=0$ though. In fact, $\delta_j H^2=\frac{\imath}{2}[\nabla_{j+1}H,\nabla_{j+2}H]$. Derivatives of higher powers can be calculated iteratively and expressed in terms of $\nabla_{j+1}H$, $\nabla_{j+2}H$ and functions of $H$.

\vspace{.2cm}

Next let us furnish $C^1_c(\Xi\times\Omega\times\ZM^d,\mbox{\rm Mat}(L,\CM))$ with the norm
$$
\| A\|_{C^1} 
\;=\; 
\|A\|\,+\, \sum_{j=1}^d \|\nabla_j A\|
\,+\, \sum_{j=1}^d \|\delta_j A\|
\;,
$$
which dominates the norm in $C^1_{ \rm s}(\Aa)$. The closure w.r.t. this norm is denoted by $C^1(\Aa)$. Lemma~\ref{lem-deltajs}(i) allows to check $\| AB\|_{C^1} \leq  \| A\|_{C^1}\| B\|_{C^1}$. As also $\| A^*\|_{C^1}=\| A\|_{C^1}$, $C^1(\Aa)$ is actually a Banach $*$-algebra. In particular, $C^1(\Aa)$ is invariant under holomorphic functional calculus. Thinking of Banach algebras of analytic functions, invariance under holomorphic functional calculus is the optimal result of general nature that one can expect. Here one, moreover, has the following fact.

\begin{proposi}
\label{prop-diffcalculus2}
Let $H = H^*$ be an element of $C^1(\Aa)$. Then for any compactly supported smooth function $f$,  the operator $f(H)$ belongs to $C^1(\Aa)$. 
\end{proposi}

\noindent {\bf Proof.} Let us provide two arguments. The first one is based on the Helffer-Sjorstrand formula as presented by Davies \cite{Dav}. In fact, Lemma~\ref{lem-deltajs}(ii) implies
$$
\|\delta_j\,(z-H)^{-1}\|
\;\leq\;
C\,|\Im m(z)|^{-3}
\;,
$$
so that hypothesis (H3) of \cite{Dav} holds, which then directly implies the result. For the second proof, let us use that  Lemma~\ref{lem-deltajs}(ii) shows
\begin{equation}
\label{eq-derivbound}
\|\delta_j\,e^{\imath t H}\|
\;\leq\;
C\,(t+t^2)
\;.
\end{equation}
Moreover, one has $\|\nabla_j\,e^{\imath t H}\|\leq Ct$ so that the bound \eqref{eq-derivbound} actually also holds in the $C^1$-norm. Now one has 
\begin{equation}
\label{eq-Fourier}
f(H)
\;=\;
\int dt\;\hat{f}(t)\;e^{\imath t H}\;,
\qquad
\hat{f}(t)\;=\;\int \frac{dE}{2\pi}\;e^{-\imath Et}\,f(E)
\;,
\end{equation}
where the integral is norm convergent in $\Aa$ because $\hat{f}$ is a Schwartz function. Therefore the above estimates show that $f(H)$ is in $C^1(\Aa)$.
\hfill $\Box$

\subsection{The von Neumann algebra and the Sobolev spaces}

For $p\in [1,\infty)$, the Banach space $L^{p}(\Aa_\BB,\Tt_\BB)$ is the closure of $\Aa_\BB$ under the norm $\| A\|_{L^p}=\left( \Tt_\BB (|A|^p) \right)^{1/p}$. If $\pi_{\mbox{\tiny GNS}}$ denotes the GNS representation of $\Tt$ on $L^2(\Aa_\BB,\Tt_\BB)$, $L^{\infty}(\Aa_\BB,\Tt_\BB)$ denotes the von Neumann algebra $\pi_{\mbox{\tiny GNS}}(\Aa_\BB)''$ where $''$ is the bicommutant. By a theorem of Connes \cite{Con}, $L^{\infty}(\Aa_\BB,\Tt_\BB)$ is canonically isomorphic to the von Neumann algebra of $\PM$-essentially bounded, weakly measurable and covariant families $(A_\omega)_{\omega\in\Omega}$ of operators on $\Hh$ endowed with the norm
$$
\| A\|_{L^{\infty}}\;=\;
\mbox{\rm $\PM$-}\!\esssup_{\omega\in\Omega}
\| A_\omega\|_{\Bb(\Hh)}
\mbox{ . }
$$
\noindent Consequently, the family of representations $\pi_\omega$ extends as a family of weakly measurable representations of $L^{\infty}(\Aa_\BB,\Tt_\BB)$. Moreover, the trace $\Tt_\BB$ extends to $L^{\infty}(\Aa_\BB,\Tt_\BB)$ as a normalized normal trace. Then $L^1(\Aa_\BB,\Tt_\BB)$ can be identified with the predual of $L^{\infty}(\Aa_\BB,\Tt_\BB)$, namely the linear space spanned by the set of normal states on $L^{\infty}(\Aa_\BB,\Tt_\BB)$. Furthermore, one disposes of an associated Sobolev space $H^1(\Aa_\BB,\Tt_\BB)$ which is given by the closure of $\Aa_\BB$ w.r.t. the norm
\begin{equation}
\label{eq-SobolevNorm}
\|A\|_{H^1}
\;=\;
\left(
\Tt_\BB(|A|^2)+\Tt_\BB(|\nabla\,A|^2)\right)^{\frac{1}{2}}
\;.
\end{equation}
It is a two-sided ideal in $L^{\infty}(\Aa_\BB,\Tt_\BB)$. The Sobolev space is closely linked to dynamical localization, see Section~\ref{sec-dynloc} below and \cite{BES}.

\vspace{.2cm}

\subsection{Time evolution}
\label{sec-timeevolve}

Let us first consider the Hamiltonian time evolution associated to a fixed self-adjoint element $H\in\Aa_\BB$. Then the associated Liouville operator $\Ll_H$ acting on $\Aa_\BB$ is defined by
\begin{equation}
\label{eq-LiouvilleH}
\Ll_H (A)\; =\; \imath\, [H,A]
\mbox{ . }
\end{equation}
\noindent It is a bounded $*$-derivation of $\Aa_\BB$ generating a one-parameter group of $*$-automorphisms $t\in\RM\mapsto e^{t\Ll_H}$ in $\Aa_\BB$. Then observables satisfy the Heisenberg equation $\partial_t A(t)=\imath[H,A(t)]$. There is a dual time-evolution of states on $\Aa_\BB$. Let us focus on normal states w.r.t.\ $\Tt_\BB$ represented by positive elements $\rho\in\Aa_\BB$ satisfying the normalization condition $\Tt_\BB(\rho)=1$. The state is then given by $A\in\Aa_\BB\mapsto\Tt_\BB(\rho A)$. Now the Hamiltonian time evolution of such states is governed by the Liouville equation $\partial_t \rho(t)=\imath[\rho(t),H]$, namely $\rho(t)=e^{-t\Ll_H}(\rho(0))$.

\vspace{.2cm}

For the theory of orbital polarization, one needs also to consider time-evolutions governed by a time-dependent Hamiltonian $t\in[0,T]\mapsto H(t)$ given by a continuous path of self-adjoint elements in $\Aa_\BB$. Then obeservables in $\Aa_\BB$ evolve according to the time-dependent Heisenberg equation
$$
\partial_t A(t)
\;=\;
\mathcal{L}_{H(t)}( A(t))
\;,
\qquad
\mathcal{L}_{H(t)}( A)
\;=\;
\imath\, [H(t),A]
\;,
$$
and normal states according to the Liouville equation 
\begin{equation}
\label{eq-Liouville}
\partial_t \rho(t)
\;=\;
-\,\mathcal{L}_{H(t)}( \rho(t))
\;.
\end{equation}
As to existence and uniqueness of these equations, let us focus on \eqref{eq-Liouville}. Because $H(t)$ is self-adjoint, the Liouville operator $\mathcal{L}_{-H(t)}$ is a bounded $*$-derivation. Using the Dyson series \cite[Theorem~X.69]{RS2}, one readily shows that $\mathcal{L}_{-H(t)}$ generates an evolution family  $(t,s)\in[0,T]^2\mapsto \eta_{t,s}$ of $*$-automorphisms satisfying $\eta_{t,t}=\one$ and $\eta_{t,r}\eta_{r,s}=\eta_{t,s}$ for $ s, r, t\in [0, T]$.  The unique solution  of \eqref{eq-Liouville} in $C^1([0,T],\Aa)$ with $\rho(0)= \rho_0\in\Aa$ is given by $\rho(t) = \eta_{t,0}(\rho_0)$. As the generators are bounded, the map $(t,s)\in [0,T]^2\mapsto \eta_{t,s}(A)\in\Aa_\Bb$ is differentiable. If one assumes in addition that $H(t)\in C^1_{\rm s}(\Aa)$ for all $t\in[0,T]$, then $\mathcal{L}_{H(t)}$ are bounded operators in the Banach space $C^1_{\rm s}(\Aa)$. Hence, if the map $t\mapsto H(t)$ is also continuous w.r.t. the norm in $C^1_{\rm s}(\Aa)$, then, again by inspecting the Dyson series, one sees that  the evolution family $\eta_{t,s}$ leaves $C^1_{\rm s}(\Aa)$ invariant and that also $(t,s)\in [0,T]^2\mapsto \eta_{t,s}(A)\in C^1_{\rm s}(\Aa)$ is differentiable. Two facts about the solutions of \eqref{eq-Liouville} are now of physical importance. First of all, as $\eta_{t,s}$ is a $*$-automorphism it follows that the particle density $\Tt(\rho(t))=\Tt(\eta_{t,0}(\rho_0))= \Tt(\rho_0)$ is conserved. Second of all, if $\rho_0=\rho_0^2$ is a projection, then also $\rho(t)^2 = \eta_{t,0}(\rho_0)\eta_{t,0}(\rho_0)= \eta_{t,0}(\rho_0^2)=\eta_{t,0}(\rho_0) = \rho(t)$.

\vspace{.2cm}

\subsection{Current operator and current-current correlation measure}
\label{sec-current}

Pysically, the current is the derivation of the position w.r.t.\ time. As these are all observables and time derivatives are thus given by Heisenberg's equation, it follows that $\dot{X}=\imath[\pi_{\BB,\omega}(H),X]=
\pi_{\BB,\omega}(\nabla H)$ where the second equality follows from \eqref{eq-derivrep}. Hence the current operator $J$ is defined for any $H\in C^1_{\rm s}(\Aa)$ as an element of $\Aa$: 
\begin{equation}
\label{eq-currentinA}
 {J} \;=\; {\nabla} H
\mbox{ . }
\end{equation}
\noindent The following result implies, in particular, that the current vanishes in all states which are functions of $H$.

\begin{proposi}
\label{prop-zerocurrent}
Let $H=H^*$ be in $C^1_{\rm s} (\Aa)$ and $f,g$ be the restrictions to the spectrum of $H$ of continuously differentiable function on $\RM$. Then 
\begin{equation}
\label{eq-orthogdeeffdeH}
\Tt_\BB ( g(H)\nabla f(H))\; =\;0
\mbox{ . }
\end{equation}
\end{proposi}

\noindent {\bf Proof.} As the trace $\Tt_\BB$ is invariant under $\nabla$, it follows from Leibniz rule that $0=\Tt_\BB(\nabla H^n)=n\Tt_\BB(H^{n-1}\nabla H)$ for all $n\geq 1$. Thus by density $\Tt_\BB(f(H)\nabla H)=0$ for any continuous function. From this follows in turn that $\Tt_\BB(f(H)\nabla H^m)=0$ for all $m\geq 0$. Again by density the result follows.
\hfill $\Box$

\vspace{.2cm}

Next let us recall \cite{KP,SBB} the definition and a few properties of the current-current correlation measure $m$. Whenever the current operators $\nabla H$ are in the algebra, or otherwise stated $H\in C^1_{ \rm s}(\Aa)$, then for each $\BB$ this measure is given by
$$
\Tt_\BB
\left( f(H) \nabla H g(H)\cdot \nabla H \right) 
\;=\;
\int_{\RM\times\RM}  m(dE,dE')\; f(E) \,g(E')
\;,
\qquad
f,g\in C_0(\RM)\;,
$$
where the dot denotes the scalar product $\nabla H  g(H) \cdot \nabla H=\sum_{j=1}^d\nabla_j H  g(H)  \,\nabla_j H$. Therefore $m$ is  a positive measure on $\RM^2$ with finite mass $\Tt_\BB(\nabla H\cdot \nabla H)$. Now by Cauchy-Schwarz
\[
|\Tt_\BB
\left( f(H) \nabla_i H g(H)\, \nabla_j H \right) |^2\;\leq\; \Tt_\BB
\left( |f|^2(H) \,\nabla_i H \,|g|^2(H)\, \nabla_i H \right)\,\Tt_\BB
\left(  \nabla_j H \,\nabla_j H \right)  
\]
and thus, by
the Radon-Nykodym theorem, there exist functions $D_{i,j}\in L^1(\RM^2,m)$ such that
$$
\Tt_\BB
\left( f(H) \nabla_i H g(H)\, \nabla_j H \right) 
\;=\;
\int_{\RM\times\RM}  m(dE,dE')\,D_{i,j}(E,E')\; f(E) \,g(E')
\;.
$$
Let us set $m_{i,j}=D_{i,j} m$. Only the diagonal measures $m_{j,j}$ are positive, but the matrix-valued measure $(m_{i,j})_{i,j=1,\ldots,d}$ is again positive. Furthermore, if $\Delta\times\Delta'$ is a rectangle, then again by Cauchy-Schwarz
$$
m(\Delta\times\Delta')
\;\leq\;
\left(
\Tt_\BB\bigl((P_{\Delta}\nabla H)^*P_{\Delta}\nabla H\bigr)
\;
\Tt_\BB\bigl((P_{\Delta'}\nabla H)^*P_{\Delta'}\nabla H\bigr)
\right)^{\frac{1}{2}}
\;\leq\;
\left(
\Tt_\BB(P_{\Delta})
\;
\Tt_\BB(P_{\Delta'})
\right)^{\frac{1}{2}}
\;\|\nabla H\|^2
\;.
$$
Therefore, if the density of states measure $\Nn$ is defined by
$$
\Tt_\BB
\left( f(H)\right) 
\;=\;
\int_{\RM}  \Nn(dE)\; f(E) 
\;,
\qquad
f\in C_0(\RM)\;,
$$
then
$$
m(\Delta\times\Delta')
\;\leq\;
C\;
\left(
\Nn(\Delta)
\,\Nn(\Delta')
\right)^{\frac{1}{2}}
\;.
$$
This shows $\alpha$-H\"older continuity of $m$ if $\Nn$ is $2\alpha$-H\"older continuous. The current-current measure allows to write out formulas for the conductivity in the relaxation time approximation and its behavior on the diagonal $E=E'$ is linked to dynamical properties of $H$ which allows to prove the anomalous Drude formula \cite{SBB}. Just below in Section~\ref{sec-dynloc}, the link of $m$ with dynamical localization will be explained. 

\vspace{.2cm}

\subsection{Dynamical localization}
\label{sec-dynloc}

This section briefly resumes the notion of dynamical Anderson localization. This is a property of random Hamiltonians $\pi_{\BB,\omega}(H)$ at a fixed magnetic field $\BB$. While there are numerous characterizations of this localized phase, here we will merely use the notion of dynamical localization in spectral (Borel) subset $\Delta\subset\RM$ in the sense that the localization length defined by
$$
\ell^2(\Delta)
\;=\;
\sup_{T>0}\;\int^T_0\frac{dt}{T}\;
\EE\;\TR\;\langle 0|\chi_\Delta(\pi_{\BB,\omega}(H))(e^{\imath \pi_{\BB,\omega}(H) t}\,X\,e^{-\imath \pi_{\BB,\omega}(H) t}-X)^2
\chi_\Delta(\pi_{\BB,\omega}(H))|0\rangle
$$
is finite. For quantum Hall systems, this is a sufficient condition for the existence of plateaux \cite{BES} and it will also emerge below as natural condition for the study of magnetization. No attempt is made here to prove it in concrete models, but it is known to hold in a strong disorder or band edge regime (see \cite{BES} for some concrete verification using the fractional moment method). It will be useful that the dynamical localization condition $\ell^2(\Delta)<\infty$ can also be checked on the current-current correlation measure, namely one has
$$
\ell^2(\Delta)
\;=\;
2\;\int_{\Delta\times\RM} m(dE,dE')\;
\frac{1}{|E-E'|^2}
\;.
$$
Now by \cite[Section~5.2]{BES}, $\ell^2(\Delta)<\infty$ implies that the Fermi projection $P_\mu=f_{\infty,\mu}(H)$ is in $H^1(\Aa_\BB,\Tt_\BB)$ for every $\mu\in\Delta$ (and depends continuously on $\mu$). Furthermore, there exists a $\Nn$-integrable function $E\in\Delta\mapsto \ell^2(E)$ such that
$$
\ell^2(\Delta')
\;=\;
\int_{\Delta'} \Nn(dE)\;\ell^2(E)\;,\qquad \mbox{ for all Borel sets}\; \Delta'\subset\Delta
\;.
$$
%

\subsection{The time reversal operator}

Time reversal is implemented in $\Aa$ by the involutive anti-automorphism $\eta:\Aa\to\Aa$ defined by
$$
(\eta(A))(\BB,\omega,n)
\;=\;
I^{-1}\,\overline{A(-\BB,\omega,n)}\,I
\;,
$$
where $I=e^{\imath\pi s^y}$ is the rotation is spin space by $180$ degree (which is unitary) and the overline denotes the complex conjugation. Note that $\eta^2=\one$, $\eta(\lambda A)=\overline{\lambda}\eta(A)$ and $I^2=\pm\one$ pending on whether the spin is even or odd. Thus one has
\begin{equation}
\label{eq-TRIalg}
\Tt_\BB(\eta(A))
\;=\;
\overline{\Tt_{-\BB}(A)}\;,
\qquad
\eta(\nabla A)\;=\;-\,\nabla (\eta(A))\;,
\qquad
\eta(\delta A)\;=\;-\,\delta (\eta(A))
\;.
\end{equation}

\section{Orbital polarization}
\label{sec-Polar}

Following up on the discussion in the introduction, this section presents the main mathematical results on the orbital polarization. Let be given a time-dependent Hamiltonian $t\in[0,T]\mapsto H(t)\in C^1_{\rm s}(\Aa)$ at fixed magnetic field $\BB$, namely a continuous path of self-adjoint elements in $\Aa_\BB$. Then states evolve according to the Liouville equation \eqref{eq-Liouville}. The change $\Delta P$ in polarization during a change of the system between time $t=0$ and $t=T$ is now defined by \eqref{eq-polardef}. It follows from Propoisition~\ref{prop-zerocurrent} that the average current density $\Tt_\BB ( \rho \,J)$ vanishes for any equilibrium state $\rho=f(H)$. However, changes of the Hamiltonian in time can give rise to non-vanishing currents so that $\Delta P$ does not necessarily vanish. Our first result is a formula for the current density solely in terms of the time-evolved state $\rho(t)$, assuming that $\rho(0)=P$ is a projection in $C^1_{\rm s}(\Aa)$.

\begin{proposi}
\label{prop-polarid}
Let $t\mapsto H(t)$ be a continuous path in $C^1_{\rm s}(\Aa)$ and
let $P\in C^1_{ \rm s}(\Aa)$  be any projection. Then the current associated to  the solution  $P(t)= \eta_{t,0}( P)$  of the Liouville equation {\rm \eqref{eq-Liouville}}   can be written as 
$$
\Tt_\BB\bigl(
P(t)\,\nabla H(t)\bigl)
\;=\;
\imath\,
\Tt_\BB\bigl(
P(t)\,[\partial_t P(t), \nabla P(t)]\bigl)
\;.
$$
%
%
\end{proposi}

\noindent {\bf Proof.} Recall that $P(t)\in C^1_{\rm s}(\Aa)$ for all $t$, so the expression on the right hand side makes sense. For sake of notational simplicity, we suppress all $t$-dependencies in the following computation.  Starting from the right hand side, one finds
\begin{eqnarray*}
\imath\;
\Tt_\BB\bigl(
P\,[\partial_t P, \nabla P]\bigl)
& = &
\Tt_\BB\bigl(
P[H,P]\nabla P
\bigl)
\;-\;
\Tt_\BB\bigl(P\,
(\nabla P)
[H,P]\bigl)
\\
& = &
-\;\Tt_\BB\bigl(
(\nabla P)
\,PH\bigl)
\;-\;
\Tt_\BB\bigl(
(\nabla P)
 \,HP\bigl)
\\
& = &
\Tt_\BB\bigl(
P\,\nabla(PH)\bigl)
\;-\;
\Tt_\BB\bigl( P
(\nabla P)
 \,H\bigl)
\\
& = &
\Tt_\BB\bigl(
P\,\nabla H \bigl)
\;.
\end{eqnarray*}
Here the first equality uses that $P(t)$ solves the Liouville equation {\rm \eqref{eq-Liouville}} , the second equality follows from  the identity $P(\nabla P)P=0$ that holds for any projection $P$, the third one is integration by parts from Lemma~\ref{lem-trace} (vi) and the final one 
is the Leibniz rule. 
\hfill $\Box$

\vspace{.2cm}

Physically the relevant initial state is the Fermi projection.  By Proposition~\ref{prop-diffcalculus} all spectral projections of $H$ that belong to a part of the spectrum separated by a gap from the rest belong to $C^1_{\rm s}(\Aa)$. Hence, Proposition~\ref{prop-polarid} is applicable if initially  the Fermi level $\mu$ lies in a gap of the spectrum of $H(0)$.

\begin{coro}
\label{cor-pola}
Assume that  the Fermi level $\mu$ lies in a gap of the spectrum of $H(0)$ and let $\rho_0 = \chi(H(0)-\mu)$, where $\chi$ denotes the characteristic function on the negative numbers. Then 
\begin{equation}
\label{eq-KV1}
\Delta P\;=\;
\imath
\int^T_0 dt\;
\Tt_\BB\bigl(
\rho(t)\,[\partial_t \rho(t), \nabla \rho(t)]\bigl)
\;.
\end{equation}
\end{coro}

The formula \eqref{eq-KV1} holds quite  generally, but it requires knowledge of the solution of the Liouville equation. In order to  express the polarization $\Delta P$ in terms of the spectral projections of $H(t)$, let us consider  \eqref{eq-Liouville} in the adiabatic limit, namely the Liouville equation
\begin{equation}
\label{eq-LiouvilleAdi}
\varepsilon\,\partial_t \rho(t)
\;=\;
\imath\,[\rho(t),H(t)]
\;,
\end{equation}
with a small adiabatic parameter $\varepsilon\ll1$. If one further supposes that the gap at the initial Fermi level $\mu(0)$ remains open  for all times $t\in[0,T]$, {\it i.e.} that there is a continuous function $\mu(t)$ with $\mu(t)\notin\sigma(H(t))$ for all $t\in[0,T]$, then the adiabatic theorem implies 
$$
\rho(t)\;=\;P_0(t)
\;+\;
\Oo(\varepsilon)
\;,
\qquad
P_0(t)\;=\;
\chi(H(t)-\mu(t))
\;.
$$
At least formally, this immediately yields a generalization of the King-Smith and Vanderbilt formula \eqref{eq-KV00} with errors of order $\Oo(\varepsilon)$. However, using superadiabatic approximations and a homotopy argument, a much stronger statement can actually be proved. Note that plugging the adiabatic approximation directly into the formula for the current yields the wrong result, since in the presence of a gap $P_0(t)=f(H(t))$ for a suitable smooth continuation $f$ of the characteristic function and thus, according to Proposition~\ref{prop-zerocurrent}, one has $\Tt(P_0(t)\nabla H)=0$. 

\begin{theo}
\label{theo-polar}
Let

\vspace{.1cm}

\noindent {\rm (i)} either $H\in C^{N+2}([0,T],C^1_{\rm s}(\Aa))$ with
$$
\partial_t^n\; H(t)|_{t=0} \;=\; \partial_t^n\; H(t)|_{t=T}\;  =\; 0\;,\qquad\mbox{ for all } 1\leq n\leq N\;,
$$

\vspace{.1cm}

\noindent {\rm (ii)} or $H\in C^{N+2}(\RM,C^1_{\rm s}(\Aa))$ be $T$-periodic.

\vspace{.1cm}
 
\noindent Further suppose that there is a continuous function $\mu(t)$ with $\mu(t)\notin \sigma(H(t))$ for all $t\in[0,T]$. Let $\rho_0 =  \chi(H(0)-\mu(0))$ in case {\rm (i)} and $\rho_0 =  P_N^\varepsilon(0)$ in case {\rm (ii)}, where $P_N^\varepsilon(0)$ is a superadiabatic projection constructed in the proof. Then
\begin{equation}
\label{eq-KV2}
\Delta P\;=\;\imath
\int^T_0 dt\;
\Tt_\BB\bigl(
P_0(t)\,[\partial_t P_0(t), \nabla P_0(t)]\bigl)
\;+\;
\Oo(\varepsilon^N)
\;.
\end{equation}
\end{theo}

\noindent {\bf Proof.}
Theorem~\ref{theo-adiabatics} in the appendix provides superadiabatic projections $P^\varepsilon_N(  t)$ which according to Corollary~\ref{coro-adiabatic}  approximate  the solution $\rho(t)$ of \eqref{eq-LiouvilleAdi} up to small errors in the sense that
$$
\| P^\varepsilon_N( t) - \rho(t) \|\;+\;\| \nabla (P^\varepsilon_N( t) - \rho(t)) \|\;+\;\| \partial_t (P^\varepsilon_N( t) - \rho(t) )\|\;= \; \Oo(\varepsilon^N)\,.
$$
With \eqref{eq-KV1} and Lemma~\ref{lem-trace}(iv) this yields that 
$$
\Delta P\;=\;\imath
\int^T_0 dt\;
\Tt_\BB\bigl(
P^\varepsilon_N( t)\,[\partial_t P^\varepsilon_N( t), \nabla P^\varepsilon_N( t)]\bigl)
\;+\;
\Oo(\varepsilon^N)
\;.
$$
But by the following calculation,  the above integral is independent of $\varepsilon$ so that one can pass to the limit $\varepsilon\to 0$ using \eqref{ST1} without changing its value. For case (i) let us use  that   the vanishing of the first $N$ derivatives of $t\mapsto H(t)$ at the endpoints implies $P^\varepsilon_N( 0) = P_0(0)$ and  $P^\varepsilon_N( T) = P_0(T)$. With the notation $\dot P_N^\varepsilon=\partial_tP^\varepsilon_N$, one finds
\begin{eqnarray*}\lefteqn{
\partial_\varepsilon\int_0^T  d t\, \Tt_\BB\left( P^\varepsilon_N[\dot P^\varepsilon_N, \nabla P^\varepsilon_N]\right) 
\;=\;
\int_0^Tdt\,  \Tt_\BB \left( P^\varepsilon_N[\partial_\varepsilon\dot P^\varepsilon_N, \nabla P^\varepsilon_N] + P^\varepsilon_N[\dot P^\varepsilon_N, \partial_\varepsilon \nabla P^\varepsilon_N]\right) }\\
&=& \Tt_\BB \left(P^\varepsilon_N[ \partial_\varepsilon P^\varepsilon_N, \nabla P^\varepsilon_N]\right)\big|_0^T - \int_0^T dt\,  \Tt_\BB \left( P^\varepsilon_N[\partial_\varepsilon P^\varepsilon_N, \nabla\dot P^\varepsilon_N]\right)  + 
\int_0^T  dt\,\Tt_\BB \left(  P^\varepsilon_N[\dot P^\varepsilon_N, \nabla\partial_\varepsilon P^\varepsilon_N]\right) \;=\;0 \,.
\end{eqnarray*}
In both equalities it was used that all the  derivatives $\dot P^\varepsilon_N$, $\partial_\varepsilon P^\varepsilon_N$ and $\nabla P^\varepsilon_N$ of the projection $P^\varepsilon_N$ are off-diagonal with respect to $P^\varepsilon_N$ and thus $\Tt_\BB( \partial_\varepsilon P^\varepsilon_N \dot P^\varepsilon_N  \nabla P^\varepsilon_N )=0$. In the second equality we used that differentiability of the map 
$\varepsilon\mapsto P_N^\varepsilon$ from $[0,\varepsilon_N)$ to $C^1([0,T], C^1_{\rm s}(\Aa))$
implies existence and equality of the mixed derivatives, $\partial_\varepsilon\dot P^\varepsilon_N = \partial_t \partial_\varepsilon P^\varepsilon_N$ and $\partial_\varepsilon\nabla P^\varepsilon_N = \nabla\partial_\varepsilon P^\varepsilon_N$.
 The last equality follows because the boundary terms vanish or cancel each other (in cases (i) and (ii) respectively) and the integrands in the last two integrals cancel each other as shows a partial integration w.r.t. $\nabla$.
 \hfill$\Box$

\vspace{.2cm}

Furthermore, the leading order term of \eqref{eq-KV2} is invariant under diffeotopies. A differentiable path $P_1:[0,T]\to C^1_{\rm s}(\Aa)$  of projections is called diffeotopic to $P_0$ whenever there exists a differentiable map $F:[0,1]\to C^1([0,T], C^1_{\rm s}(\Aa))$ such that   $F(0,t) = P_0(t)$ and $F(1,t) = P_1(t)$. Replacing $\varepsilon$ in the proof of Theorem~\ref{theo-polar} by the diffeotopy parameter, one obtains the following.

\begin{coro}\label{coro-diff}
Let either of the hypothesis in {\rm Theorem~\ref{theo-polar}} hold and suppose that the diffeotopy between the paths $P_0$ and $P_1$ satisfies, in case {\rm (i)}  $F(\alpha,0) = P_0(0)$ and $F(\alpha,T) = P_0(T)$ for all $\alpha\in [0,1]$, or, in case {\rm (ii)}, $F(\alpha,0) =F(\alpha,T)$ for all $\alpha\in [0,1]$. Then 
$$
\int^T_0 dt\;
\Tt_\BB\bigl(
P_0(t)\,[\partial_t P_0(t), \nabla P_0(t)]\bigl) \;=\;
\int^T_0 dt\;
\Tt_\BB\bigl(
P_1(t)\,[\partial_t P_1(t), \nabla P_1(t)]\bigl)\,.
$$
\end{coro}

\vspace{.2cm}
This explains the experimental fact that the total polarization depends only on the initial and final configuration of the solid. However, this only holds if two paths are diffeotopic. For general cyclic deformations, {\it i.e.}  $H(T)=H(0)$ and thus $P_0(T)=P_0(0)$, the super-adiabatic approximation of the charge transported  in one cycle
$$
 \int^T_0 dt\;
\Tt_\BB\bigl(
P_0(t)\,[\partial_t P_0(t), \nabla P_0(t)]\bigl) \;\;\in\;\;2\,\pi\, \mathbb{Z}^d
$$
is actually a topological quantity, as was already emphasized by Thouless \cite{Tho} in a more restricted context. To see this, let us consider the C$^*$-algebra $\widehat{\Aa}=C(\SM^1_T)\otimes\Aa$ of continuous functions on the circle $\SM^1_T\cong[0,T)$ with values in $\Aa$. Operators therein will be denoted by $\widehat{A}=(A_{t})_{t\in\SM^1_T}$ with $A(t)\in\Aa$. Besides the derivations $\nabla$, this algebra has the densely defined $*$-derivation $\imath\,\partial_t$ and a  trace
$$
\widehat{\Tt}_\BB(\widehat{A})
\;=\;
\int^T_0  dt \;
\Tt_\BB(A(t))
\;.
$$
Thus the polarization for a $T$-periodic Hamiltonian is given by
$$
\Delta P
\;=\;
2\,\pi\,  \mbox{\rm Ch}(\widehat{P}_0)\;+\;\Oo(\varepsilon^N)
\;,
$$
where $\widehat{P}_0=(P_0(t))_{t\in\SM^1_T}\in\widehat{\Aa}$ and the vector of associated Chern numbers of a differentiable $\widehat{P}\in\widehat{\Aa}$   defined by
$$
\mbox{\rm Ch}(\widehat{P})
\;=\;
\frac{1}{2\pi\imath}\;
\widehat{\Tt}_\BB
\bigl(\widehat{P}\,[\imath\,\partial_t \widehat{P},\nabla \widehat{P}]\bigr)
$$
has integer components.
This shows that even in the presence of disorder the charge transported  under cyclic deformations is quantized.  
  
\vspace{.2cm}

\begin{coro}
\label{coro-polar}
Let $H_\lambda(t) = H_0(t) + \lambda V(t)$ be such that   $H_\lambda(t)$ satisfies the assumptions of {\rm Theorem~\ref{theo-polar}, case (ii)}, for $\lambda \in [0,1]$ with a continuous choice for $\mu(t,\lambda)$. Then the charge transported in a periodic cycle of $H_\lambda(t)$ is quantized and agrees with that of $H_0(t)$, up to errors of order $\varepsilon^N$.
\end{coro}

\noindent{\bf Proof.}
It follows from standard perturbation arguments, {\it c.f.} Lemma~\ref{lemm-pert} in the appendix, that the Fermi projections of $H_\lambda(t)$ and $H_0(t)$ are diffeotopic in the sense of Theorem~\ref{theo-polar}.
 \hfill$\Box$

\section{Magnetization at finite temperature}
\label{sec-magn}

In this section, the magnetization at finite temperature will be calculated for Hamiltonians in dimension $d=2$ or $d=3$. It will be expressed in terms of the current-current correlation measure. Let us begin by recalling the definitions \eqref{eq-pressuredef} and \eqref{eq-magdef} of the pressure $p({\beta,\mu})$ and magnetization $M({\beta,\mu})$ at inverse temperature $\beta$ and chemical potential $\mu$. If $\BB=0$ and the time reversal symmetry relation $\eta(H)=H$ holds at $\BB=0$, then one deduces from Lemma~\ref{lem-deltajs}(v) and the identities \eqref{eq-TRIalg} that
$$
M({\beta,\mu})
\;=\;
\Tt_0\bigl(\delta\,
\ln(\one+e^{-\beta(\eta(H)-\mu)})
\bigr)
\;=\;
\Tt_0\bigl(\delta\,\eta(
\ln(\one+e^{-\beta(H-\mu)}))
\bigr)
\;=\;
-\,\overline{M({\beta,\mu})}
\;.
$$
As the magnetization is real, it follows that it vanishes for any time reversal symmetric system so that we suppose from now on time reversal symmetry to be broken. This does {\it not} necessarily mean that $\BB\not = 0$ though as there may be local magnetic fields breaking the time reversal symmetry. In order to calculate the magnetization, let us use the identity
$$
\ln(1+e^{-\beta E})
\;=\;
\ln(2)+\int^\beta_0 d\beta'
\;\partial_{\beta'}\,
\ln(1+e^{-\beta' E})
\;=\;
\ln(2)+\int^\beta_0 d\beta'
\;
(1+e^{\beta' E})^{-1}\, (-E)
\;.
$$
Thus, if $f_{\beta,\mu}(H)=(1+e^{\beta(H-\mu)})^{-1}$ is the Fermi-Dirac distribution, the cyclicity of the trace shows
\begin{equation}
\label{eq-Magint}
M_j({\beta,\mu})
\;=\;
\frac{1}{\beta}\;\int^\beta_0 d\beta'
\;
\partial_{B_j}\,
\Tt_\BB
\bigl(f_{\beta',\mu}(H)\,(\mu-H)\bigr)
\;.
\end{equation}
Thus it is convenient to set
$$
\widetilde{M}_j(\beta,\mu)
\;=\;\partial_{B_j}\,\Tt_\BB\bigl(f_{\beta,\mu}(H)\,(\mu-H)\bigr)
\;.
$$
The next result follows from the observation that $\widetilde{M}_j(\beta,\mu)$ can be calculated in terms of the current-current correlation measure $m$ defined in Section~\ref{sec-current}.

\begin{theo}
\label{theo-magT}
Let  $\delta H=0$. Then, with $j$ taken cyclically,  
\begin{equation}
\label{eq-magT}
{M}_j({\beta,\mu})
\;=\;
\imath
\int_{\RM^2}m_{j+1,j+2}(dE,dE')
\;g_{\beta,\mu}(E,E')
\;,
\end{equation}
where $g_{\beta,\mu}$ is the following smooth function
\begin{equation}
\label{eq-gsplitting}
g_{\beta,\mu}(E,E')
\;=\;\frac{1}{2}\,
\frac{1}{E'-E}\left(f_{\beta,\mu}(E)
\;+\;
\frac{1}{\beta}\;
\frac{\ln(1+e^{-\beta (E'-\mu)})- \ln( 1+e^{-\beta (E-\mu)})}{E'-E}
\right)
\;-\;
(E\leftrightarrow E')
\;.
\end{equation}
\end{theo}

\noindent {\bf Proof.} In order to simplify the formulas, let us shift $H$ to $H-\mu$ so that we may assume that $\mu=0$ and suppress the $\mu$-dependence in the following. Let us also choose a smooth compactly supported function $f$ that is equal to $f_{\beta}$ on the spectrum of $H$. Its Fourier transform  with choice of normalization as in \eqref{eq-Fourier} is denoted by $\hat{f}$. Let us also set $F=f(H)$. Then one has
$$
\widetilde{M}_j({\beta})
\;=\;
-\,\Tt_\BB(H\delta_j F)
\;=\;
-\,\int dt\;\hat{f}(t)\,\Tt_\BB(H\delta_j e^{\imath Ht})
\;.
$$
Now replacing the generalized DuHamel formula, using the cyclicity of the trace and replacing the current-current correlation measure shows
\begin{eqnarray*}
\widetilde{M}_j({\beta})
& = &
-\,\int dt\,\hat{f}(t)\int^1_0 ds\int^s_0 dr\;\frac{\imath}{2}\,(\imath t)^2
\,
\Tt_\BB
\left(
H\,[e^{\imath Ht(1-r)}\nabla_{j+1}H,e^{\imath Htr}\nabla_{j+2}H]
\right)
\\
& = &
\int dt\,\hat{f}(t)\int^1_0 dr\,(1-r) \;\frac{\imath}{2}\,t^2
\int m_{j+1,j+2}(dE,dE')\;
\left(Ee^{\imath Et(1-r)}e^{\imath E'tr}
\;-\;E'e^{\imath E't(1-r)}e^{\imath Etr}
\right)
\\
& = &
\int dt\,\hat{f}(t)\;\frac{\imath}{2}\,t^2
\int m_{j+1,j+2}(dE,dE')\;
\left(
\frac{Ee^{\imath Et}+E'e^{\imath E't}}{\imath(E-E')t}
\;+\;
\frac{(E+E')(e^{\imath Et}-e^{\imath E't})}{(E-E')^2t^2}
\right)
\\
& = &
\frac{\imath}{2}\,
\int m_{j+1,j+2}(dE,dE')\;
\left(
\frac{Ef'(E)+E'f'(E')}{E'-E}
\;-\;
\frac{(E+E')(f(E')-f(E))}{(E'-E)^2}
\right)
\;.
\end{eqnarray*}
Now let us replace $\widetilde{M}_j({\beta})$ into \eqref{eq-magT} and evaluate the integral over $\beta'$. For that purpose one can again use $f=f_\beta$. The following identities are then useful:
$$
\int^\beta_0 d\beta'\;f_{\beta'}(E)
\;=\;
-\,\frac{1}{E}\,\ln(1+e^{-\beta E})
\;,
\qquad
\int^\beta_0 d\beta'\;f'_{\beta'}(E)
\;=\;
\frac{1}{E^2}\,\ln(1+e^{-\beta E}) \;+\;\frac{\beta}{E}\,f_\beta(E)
\;.
$$
After a short algebraic calculation, one finds
\begin{equation}
\label{eq-gsplit}
{M}_j({\beta})
\;=\;
\frac{\imath}{2}\,
\int m_{j+1,j+2}(dE,dE')\;
\left(
\frac{f(E)+f(E')}{E'-E}
\;+\;
\frac{2}{\beta}\;\frac{\ln(1+e^{-\beta E'})- \ln( 1+e^{-\beta E})}{(E'-E)^2}
\right)
\;.
\end{equation}
Regrouping terms and shifting energy back by $\mu$ allows to conclude the proof. The smoothness of $g_{\beta,\mu}$, in particular, on the diagonal $E=E'$ is readily checked.
\hfill $\Box$

\vspace{.2cm}

As an illustration, let us consider the case of a $1$-periodic operator at zero magnetic field. Then Bloch theory applies and $\Aa_{\BB=0}=C(\TM^d)\otimes\mbox{\rm Mat}(L\times L,\CM)$. Moreover, $\Tt_{\BB=0}(A)=\int_{\TM^d} \tfrac{d k}{(2\pi)^d} \,\TR(A(k)) $ and $\nabla_j A(k)=\partial_j A(k)$ for a differentiable $A$ where $\partial_j$ denotes the partial derivative w.r.t. $k_j$. The Hamiltonian $H\in C^1(\TM^d)\otimes\mbox{\rm Mat}(L\times L,\CM)$ is given by the matrix-valued function $k\in \TM^d\mapsto H(k)$. Let $E_1(k)\leq\ldots\leq E_L(k)$ be the corresponding eigenvalues, namely the band functions, and let $P_l(k)$  be the eigenprojection corresponding to $E_l(k)$. In this situation, the magnetization can readily be evaluated and gives the following formula stated in \cite{XYFN,SVXN}. The limit $\beta\to\infty$ can be taken and this then also shows formula (55) in \cite{Res}.

\begin{coro}
\label{coro-magTperiodic}
Let $H$ be $1$-periodic, $\delta H=0$ and $\BB=0$. Suppose that all bands are isolated, i.e.\  $E_1(k)<E_2(k)<\ldots< E_L(k)$ for all $k\in \TM^d$. Further let the Rammal-Wilkinson tensor $R^{(l)}(k)$ and curvature of the Berry connection $\Omega^{(l)}(k)$ for the $l$th Bloch band be given by
$$
R^{(l)}_{i,j}(k)
\;=\; \frac{\imath}{2}\Big(
\TR \bigl(P_l(k)\,\partial_i P_l(k)\,(H(k)-E_l(k))\,\partial_jP_l(k)\bigr)-
\TR \bigl(P_l(k)\,\partial_j P_l(k)\,(H(k)-E_l(k))\,\partial_iP_l(k)\bigr)\Big)
\;,
$$
and 
$$
\Omega^{(l)}_{i,j}(k)
\;=\; \imath\;
\TR \bigl(P_l(k)\,[\partial_i P_l(k), \partial_jP_l(k)]\bigr)
\;.
$$
Then the magnetization is given by
\begin{equation}\label{Bloch-magni}
{M}_j({\beta},\mu) 
\;=\; 
\sum_{l=1}^L  \int\frac{d k}{(2\pi)^d} \left(
f_{\beta,\mu}(E_l(k)) \,R^{(l)}_{j+1,j+2}(k)\, +\, 
\frac{1}{\beta}\, \ln\left(1+ e^{-\beta (E_l(k)-\mu) }\right) \Omega^{(l)}_{j+1,j+2}(k)
\right)\,.
\end{equation}
\end{coro}

\noindent {\bf Proof.}  Then the current-current correlation measure is given by
\begin{eqnarray*}
\int m_{i,j}(dE,dE')\,f(E)g(E')
& = &
 \sum_{l=1}^L
\int\frac{d k}{(2\pi)^d}\;
f(E_l(k)) \,g(E_l(k))\;
\partial_{i} E_l(k)\partial_{j} E_l(k)
\\
& + & \!\!\!\!
 \sum_{l,n=1,\, l\not=n}^L
\int\!\!\frac{d k}{(2\pi)^d}\,
f(E_l(k)) \,g(E_{n}(k))\,
\TR\bigl(P_l(k) \partial_{i}H(k) P_{n}(k) \partial_{j}H(k)\bigr)
\;.
\end{eqnarray*}
This is replaced into the formula \eqref{eq-magT} for ${M}_j(\beta,\mu)$. As
$$
\lim_{E'\to E} \;g_{\beta,\mu}(E,E')
\;=\;
\frac{1}{4}\,f'_{\beta,\mu}(E)\,-\,\frac{1}{4}\,f'_{\beta,\mu}(E)
\;=\;0
\;,
$$
the diagonal term vanishes and one finds
\begin{equation}
\label{eq-Magperiodic}
{M}_j(\beta,\mu) \; = \;\frac{\imath}{2}\;
\sum_{l,n=1,\, l\not=n}^L
   \int\frac{d k}{(2\pi)^d}\;
g_{\beta,\mu}(E_l(k),E_{n}(k))
\; \TR\bigl(P_l(k) \partial_{j+1}H(k) P_n(k) \partial_{j+2}H(k)\bigr)\;.
\end{equation}
This can be further simplified. To make formulas more simple, we suppress the dependence of all objects on $k$ in the following formulas. Deriving $P_lHP_n=0$ and $P_lP_n=0$, one obtains the identities
$$
0 
\;=\; P_l \,\partial_j H \,P_n \;+\; \partial_j P_l \,E_n\,P_n \;+\;  P_l\,E_l \,\partial_j P_n
\;,
\qquad
0 \;=\; \partial_j P_l\, P_n \;+\;  P_l\,\partial_j P_n
\;,
$$
from which a straightforward computation shows 
$$
\TR\bigl(P_l \, \partial_{j }H  \,P_{n } \, \partial_{i}H \bigr) = (E_l-E_n)^2 \;\TR \bigl(P_l\,\partial_j P_l\,P_n\,\partial_iP_l\bigr)\,,
$$
Therefore, in the splitting of $g_{\beta,\mu}$ as in \eqref{eq-gsplit}, the first term in \eqref{eq-Magperiodic} simplifies to 
$$
\frac{\imath}{2}\,\sum_{l,n=1,\, l\not=n}^L
\frac{f_{\beta,\mu}(E_l) + f_{\beta,\mu}(E_n)}{E_n-E_l} (E_l-E_n)^2 \;\TR \bigl(P_l\,\partial_{j+1} P_l\,P_n\,\partial_{j+2}P_l\bigr)
\;=\;  
\sum_{l=1}^L f_{\beta,\mu}(E_l) \, R^{(l)}_{j+1,j+2}
\;,
$$
and the second to
$$
\frac{ \imath}{\beta}\,
\sum_{l,n=1,\, l\not=n}^L
\!\!\left(\ln(1+ e^{-\beta E_l}) - \ln(1+ e^{-\beta E_n})
\right)  \;\TR \bigl(P_l\,\partial_{j+1} P_l\,P_n\,\partial_{j+2}P_l\bigr)
\,=\,
\frac{1}{\beta} \sum_{l=1}^L \ln(1+ e^{-\beta E_l})\, \Omega^{(l)}_{j+1,j+2}
\;.
$$
The sum of these two thus leads to  formula (\ref{Bloch-magni}).
\hfill $\Box$

\vspace{.2cm}

The derivation of (\ref{Bloch-magni}) in  \cite{XYFN} is based on the semi-classical model for periodic solids including first order corrections. A rigorous justification of this approach   is the content of \cite{ST}.

\section{Magnetization at zero temperature}
\label{sec-mag0}

In this section we will focus on the case of zero temperature, namely $\beta\to\infty$. 
Due to \eqref{eq-Magint}, one  has 
\begin{equation}
\label{eq-Mequality}
M_j({\infty,\mu})
\;=\;
 \lim_{\beta\to \infty} \widetilde{M}_j(\beta,\mu)
\;,
\end{equation}
whenever the limit   exists. Even though a stronger result is proved in Theorem~\ref{theo-magT=0} below, let us first present a calculation of $\widetilde{M}_j(\infty,\mu)= \partial_{B_j}\,\Tt_\BB\bigl((\mu-H)P_\mu\bigr)$ in the presence of a spectral gap at the Fermi level $\mu$. This serves as motivation for the following and, moreover, the argument is elucidating. 

\begin{proposi}
\label{prop-magT=0}
Suppose that the chemical potential $\mu$ lies in a gap of the spectrum of $H$.  Let $P_\mu=f_{\infty,\mu}(H)$ denote the Fermi projection and suppose that $\delta H=0$. Then, with $j$ taken cyclically,
\begin{eqnarray}
\label{eq-magT=0}
\widetilde{M}_j({\infty,\mu})
& = &
\frac{\imath}{2}\;\Tt_{\BB}\bigl((\mu-H)(\one-2P_\mu)[\nabla_{j+1}P_\mu,\nabla_{j+2}P_\mu]\bigr)
\\
& = &
-\,\frac{\imath}{2}\;\Tt_{\BB}\bigl(|\mu-H|\,[\nabla_{j+1}P_\mu,\nabla_{j+2}P_\mu]\bigr)
\;.
\nonumber
\end{eqnarray}
\end{proposi}

\noindent {\bf Proof.} Because under the gap condition $P=P_\mu$ is smooth, compactly supported function  of the self-adjoint element $H$ in the Banach $*$-algebra $C^1(\Aa)$, it follows from Proposition~\ref{prop-diffcalculus2} that $P\in C^1(\Aa)$. This allows us to use the identities of Lemma~\ref{lem-deltajs}. Let us start from
$$
\widetilde{M}_j(\infty,\mu)
\;=\;\partial_{B_j}\,\Tt_\BB\bigl((\mu-H)P\bigr)
\;=\;
\Tt_\BB\bigl(\delta_j ((\mu-H)P)\bigr)
\;.
$$
Now the product rule for $\delta_j$ and $\delta_j H=0$ shows
$$
\widetilde{M}_j(\infty,\beta)
\;=\;
\Tt_\BB\bigl((\mu-H)\delta_j P\bigr)
\;+\;\frac{\imath}{2}\,
\Tt_\BB\bigl(\nabla_{j+1} (\mu-H)\,\nabla_{j+2}P-\nabla_{j+2} (\mu-H)\,\nabla_{j+1}P)\bigr)
\;.
$$
But partial integration and the fact that the derivations $\nabla_j$ all commute shows for any two operators $A,B\in H^1(\Aa_\BB,\Tt_\BB)$, 
\begin{equation}
\label{eq-partint}
\Tt_\BB\bigl(\nabla_{j} A\nabla_{j'}B\bigr)
\;=\;
-\,\Tt_\BB\bigl(A\,\nabla_{j}\nabla_{j'}B\bigr)
\;=\;
-\,\Tt_\BB\bigl(A\,\nabla_{j'}\nabla_{j}B\bigr)
\;=\;
\Tt_\BB\bigl(\nabla_{j'}A\,\nabla_{j}B\bigr)
\;.
\end{equation}
Hence
$$
\widetilde{M}_j(\infty,\beta)
\;=\;
\Tt_\BB\bigl((\mu-H)\delta_j P\bigr)
\;.
$$
This is the trace of $\delta_j P$ multiplied by a function of $H$. Such a quantity can be calculated due to the following fact. If one multiplies $\delta_i P=\delta_i P^2$ calculated with the product rule of Lemma~\ref{lem-deltajs}(i) from the left and right by $P$ shows
\begin{equation}
\label{eq-diagdelta}
P\,\delta_iP\,P
\;=\;
\frac{1}{2\,\imath}\;
P\,[\nabla_{i+1}P,\nabla_{i+2}P]\,P
\;.
\end{equation}
Similarly,
\begin{equation}
\label{eq-diagdelta2}
(\one-P)\,\delta_iP\,(\one-P)
\;=\;
\frac{\imath}{2}\;
(\one-P)\,[\nabla_{i+1}P,\nabla_{i+2}P]\,(\one-P)
\;.
\end{equation}
This is to be compared with the fact that for any derivation $\nabla$ and projection $P$ one has $P\nabla PP=0$, so the r.h.s. of the last two equations are correction terms to this. Now one can replace these identities into
$$
\widetilde{M}_j(\infty,\beta)
\; = \;
\Tt_\BB\bigl((\mu-H)P\,\delta_j P\,P\bigr)
\;+\;
\Tt_\BB\bigl((\mu-H)(\one-P)\,\delta_j P\,(\one-P)\bigr)
\;,
$$
in order to conclude the proof of \eqref{eq-magT=0}, from which the second identity follows immediately.
\hfill $\Box$

\vspace{.2cm}

A natural question is now whether and under which hypothesis $\widetilde{M}_j(\beta,\mu)$ converges to the expression~\eqref{eq-magT=0} in the limit $\beta\to\infty$. If such convergence is given, then \eqref{eq-Mequality} shows that the zero-temperature magnitization itself is given by \eqref{eq-magT=0}. This will be shown to be true if there is a gap, but also under the more general hypothesis that the Fermi level $\mu$ lies in a region of dynamically localized states. Indeed, the expression~\eqref{eq-magT=0} continues to be well-defined if $P\in H^1(\Aa_\BB,\Tt_\BB)$ so that the Sobolev norm \eqref{eq-SobolevNorm} is finite (by the way, the same holds for Streda's formula and the Hall conductance). But finiteness of the Sobolev norm is known to be linked to finiteness of the dynamical localization length $\ell^2(\Delta)$  as is recalled in Section~\ref{sec-dynloc}.  If dynamical localization is given, the next result expresses the zero-temperature magnetization in terms of the current-current correlation measure. Formally, this is simply obtained by taking the limit of the integrand in \eqref{eq-magT}.

\begin{proposi}
\label{prop-magTcorrect}
Suppose that  $\delta H=0$ and the density of states has no atom at the Fermi level $\mu$. Moreover, let $\mu$ lie in an interval $\Delta$ of dynamical localization, namely $\ell^2(\Delta)<\infty$. Then
\begin{equation}
\label{eq-maginm}
\lim_{\beta\to\infty}\;
{M}_j({\beta,\mu})
\;=\;
\frac{\imath}{2}
\int_{\RM^2}m_{j+1,j+2}(dE,dE')
\;\frac{(E+E'-2\mu)(\chi(E-\mu) - \chi(E'-\mu))}{(E'-E)^2}
\;.
\end{equation}
\end{proposi}

\noindent {\bf Proof.} First of all, let us point out that the integral on the r.h.s. of the claim is well-defined and finite due to the localization hypothesis because $m_{i,j}$ is dominated by $m$. Also let us again shift energy so that $\mu=0$. Then the integrand $g_\beta$ in formula \eqref{eq-magT} for the magnetization has the following limit $g_\infty=\lim_{\beta\to\infty}\; g_{\beta}$:
$$
g_{\infty}(E,E') 
 \;=\; 
 \frac{(E+E')(\chi(E) - \chi(E'))}{(E'-E)^2}
 \,,
$$
because de l'Hopital's rule implies
$$
 \lim_{\beta\to\infty}
 \;\frac{\ln(1+e^{-\beta E})}{\beta}
 \;=\;
 -\,E\,\chi(E)
 \;.
 $$
Therefore, one has to show that the integral w.r.t. to $m_{j+1,j+2}$ of the function $h_\beta=g_\beta-g_\infty$ vanishes in the limit $\beta\to\infty$. For this purpose, let us split the integration region $\RM^2$ into three disjoint parts $R_1$, $R_2$ and $R_3$ where 
\begin{eqnarray*}
R_1 &=& \{ (E,E') \in\RM^2 \,|\, |E|<\beta^{-\eta }  \} \,\cup\,  \{ (E,E') \in\RM^2 \,|\, |E'|<\beta^{-\eta }  \} 
\;,
\\
R_2 &=&  \{ (E,E') \in\RM^2 \,|\, |E-E'|<\beta^{-N}\}\setminus R_1
\;,
\\
R_3 &=& \RM^2\setminus (R_1\cup R_2)
\;,
\end{eqnarray*}
for some $0<\eta<1$ and $N\in \NM$. On the two strips making up $R_1$, the integrand is bounded by
$$
|h_\beta(E,E')|
\;\leq\;
\frac{C}{(E-E')^2}
\;.
$$
For $\beta$ sufficiently large, the interval $I_\beta = [-\beta^{-\eta },\beta^{-\eta }]$ is contained in an interval of dynamical localization. Therefore, using that $m$ is symmetric in the variables $(E,E')$,
$$
\left|
\int_{R_1}\! m_{i,j}(dE,dE')
\;h_\beta(E,E')\;
\right|
\;\leq\;
2
\int_{\RM\times I_\beta}\!\!m(dE,dE')
\,|h_\beta(E,E')|
\;\leq\;
2\,C\,\ell^2(I_\beta)
\;=\;
\int _{I_\beta}\Nn(dE)\;\ell^2(E)
\;.
$$
Because of the continuity of the density of states measure $\Nn$, this vanishes in the limit $\beta\to\infty$ for every $\eta >0$. In the region $R_2$, the limit $g_\infty$ vanishes identically and hence   $h_\beta=g_\beta$. Using that  $f_\beta(E)= F'_\beta(E)$ is the $E$-derivative of $F_\beta(E)=-\,\tfrac{1}{\beta}\ln(1+e^{-\beta E})$, a Taylor expansion of $F$  yields
 $$
h_\beta (E,E+\delta)
\;=\; 
\frac{\delta (F'_\beta(E)+F'_\beta(E+\delta)) -2 F_\beta(E+\delta) +2F_\beta(E)}{2\,\delta^2} 
\;=\; \Oo(\delta)\,.
$$
Hence the integrand $h_\beta = g_\beta$  is uniformly of order $\Oo(\beta^{-N})$ on $R_2$. For the estimate on $R_3$, let us rewrite the estimate as
$$
h_\beta(E,E')
\; = \;
\frac{f_\beta(E)-\chi(E)+f_\beta(E')-\chi(E')}{E'-E}
\;+\;2\;
\frac{F_\beta(E)-E\chi(E)-F_\beta(E')+E'\chi(E')}{(E-E')^2}
\;,
$$
and then use that 
$$
| f_\beta(E)- \chi(E) |
\; =\; 
\Oo(e^{-\beta E})\;,
\qquad
\left|
F_\beta(E)\,-\,E\,\chi(E)\right|
 \;=\;\Oo(E e^{-\beta E}) 
\;,
$$
to conclude that for $E,E'\in R_3$, $|E|\leq\|H\|$ and $|E'|\leq\|H\|$
$$
|h_\beta(E,E')|
\;\leq\;
C\;\beta^{2N}\,e^{-\beta^{1-\eta }}
\;.
$$
Hence on the regions $R_2$ and $R_3$ the integrand is smaller than any power of $1/\beta$.
\hfill $\Box$

\vspace{.2cm}

Finally, one can show that \eqref{eq-magT=0} is indeed the correct expression for the zero temperature magnetization even in a region of dynamical localization. This confirms that the magnetization is of geometric origin.

\begin{theo}
\label{theo-magT=0}
Suppose that the chemical potential $\mu$ lies in an interval $\Delta$ with finite localization length $\ell^2(\Delta)$ and that the density of states has no atom at $\mu$.  Let $P_\mu=f_{\infty,\mu}(H)$ denote the Fermi projection and suppose that $\delta H=0$. Then, with $j$ taken cyclically,
\begin{eqnarray}
\label{eq-magT=0full}
{M}_j({\infty,\mu})
&=&
\frac{\imath}{2}\;\Tt_{\BB}\bigl((\mu-H)(\one-2P_\mu)[\nabla_{j+1}P_\mu,\nabla_{j+2}P_\mu]\bigr)
\\
&=&
-\,\frac{\imath}{2}\;\Tt_{\BB}\bigl(|\mu-H|\,[\nabla_{j+1}P_\mu,\nabla_{j+2}P_\mu]\bigr)
\;.
\end{eqnarray}
\end{theo}

\noindent {\bf Proof.} Again we set $P=P_\mu$. Let $I$ denote the r.h.s. of \eqref{eq-magT=0full}. It is split into four contributions:
\begin{eqnarray*}
I
& = &
\frac{\imath}{2}\;\Tt_{\BB}
\Bigl(
(\mu-H)(\one-P)\,\nabla_{j+1}P\, P\, \nabla_{j+2}P\,(\one-P)
-(j+1\leftrightarrow j+2)
\Bigr)
\\
& & 
\;-\,
\frac{\imath}{2}\;\Tt_{\BB}
\Bigl(
(\mu-H)P\,\nabla_{j+1}P\,(\one-P)\,\nabla_{j+2}P\,P
-(j+1\leftrightarrow j+2)
\Bigr)
\;.
\end{eqnarray*}
Focussing on the first one, it is useful to set
$$
J\;=\;
\Tt_{\BB}
\Bigl(
(\mu-H)(\one-P)\,\nabla_{j+1}P\, P\, \nabla_{j+2}P\,(\one-P)
\Bigr)
\;.
$$
Next let $Q_\eta=\chi(H-\mu-2\eta)$ for $\eta>0$. Then $PQ_\eta=0$ and $(1+P)Q_\eta=Q_\eta$. Moreover, by hypothesis
$$
\lim_{\eta\downarrow 0}\,\Tt_\BB(\one-P-Q_\eta)
\;=\;
\lim_{\eta\downarrow 0}\,\int_{[\mu,\mu+2\eta]}\Nn(dE)
\;=\;
0
\;.
$$
Therefore, using Lemma~\ref{lem-trace}(iv) one has
$$
J
\;=\;
\lim_{\eta\downarrow 0}\;
\Tt_{\BB}
\Bigl(
(\mu-H)Q_\eta\,\nabla_{j+1}P\, P\, \nabla_{j+2}P\,Q_\eta
\Bigr)
\;.
$$
Next let us use
$$
0\;=\;
\nabla_j[H,P]
\;=\;
[\nabla_jH,P]\,-\,\imath\,\Ll_H(\nabla_j P)
\;,
$$
which implies that for any $\epsilon>0$
\begin{equation}
\label{eq-beforeinvert}
(\Ll_H+\epsilon)(P\,\nabla_jP\, Q_\eta)
\;=\;
\imath\,P\,\nabla_j H\,Q_\eta
\,+\,\epsilon\,P\,\nabla_jP\, Q_\eta
\;.
\end{equation}
In order to calculate $P\,\nabla_jP Q_\eta$, one now wants to invert $\Ll_H+\epsilon$. This is indeed possible because $\Ll_H$ is an anti-self-adjoint super-operator on $L^2(\Aa,\Tt_\BB)$. Moreover, one has:

\vspace{.1cm}

\noindent {\bf Claim:} For any bounded operator $B$, the equation $\Ll_H(PA\,Q_\eta)=PB\,Q_\eta$ has  a unique bounded solution $PA\,Q_\eta=\Ll_H^{-1}(PB\,Q_\eta)$.

\vspace{.1cm}

Indeed, the equation can be written in the form
$$
P(H-\mu-\eta)(PA\,Q_\eta)-(PA\,Q_\eta)Q_\eta(H-\mu-\eta)
\;=\;
\imath\,PB\,Q_\eta\;.
$$
Now $P(H-\mu-\eta)\leq -\eta\, P$ and $Q_\eta(H-\mu-\eta)\geq \eta\, Q_\eta$. Therefore
$$
S\;=\;\int^\infty_0 dt\;e^{tP(H-\mu-\eta)}\,(\imath\, PB\,Q_\eta)\,e^{-tQ_\eta(H-\mu-\eta)}
\;,
$$
is well-defined because the integral converges. But one readily checks $\Ll_H(S)=PBQ_\eta$ so that the $S=\Ll_H^{-1}(PBQ_\eta)$. Moreover, $\Ll_H$ has a large kernel, namely all functions of $H$, but the off-diagonal part in the direct sum representation of $P$ and $\one-P$ does not contain any vector in the kernel. This proves the claim.

\vspace{.2cm}

Now inverting the operator $\Ll_H+\epsilon$ in \eqref{eq-beforeinvert} and taking the limit $\epsilon\to 0$, one finds
$$
P\,\nabla_jP\, Q_\eta
\;=\;
\imath\,(\Ll_H)^{-1}(P\,\nabla_j H\,Q_\eta)
\;.
$$
Similarly,
$$
Q_\eta\,\nabla_jP\, P
\;=\;
-\,\imath\,(\Ll_H)^{-1}(Q_\eta\,\nabla_j H\,P)
\;.
$$
Now let us replace these identities into the formula for $J$:
$$
J\;=\;
\lim_{\eta\downarrow 0}\;
\Tt_{\BB}
\Bigl(
(\mu-H)
(\Ll_H)^{-1}(Q_\eta\,\nabla_j H\,P)
(\Ll_H)^{-1}(P\,\nabla_j H\,Q_\eta)
\Bigr)
\;.
$$
Dealing in a similar manner with the three other terms, one thus finds
\begin{eqnarray*}
I
& = &
\lim_{\eta\downarrow 0}\;
\frac{\imath}{2}\;\Tt_{\BB}
\Bigl(
(\mu-H)Q_\eta\,\Ll_H^{-1}(\nabla_{j+1}H)\, P\, \Ll_H^{-1}(\nabla_{j+2}H)\,Q_\eta
-(j+1\leftrightarrow j+2)
\Bigr)
\\
& & 
\;-\,
\frac{\imath}{2}\;\Tt_{\BB}
\Bigl(
(\mu-H)P\,\Ll_H^{-1}(\nabla_{j+1}H)\,Q_\eta\,\Ll_H^{-1}(\nabla_{j+1}H)\,P
-(j+1\leftrightarrow j+2)
\Bigr)
\;.
\end{eqnarray*}
As there are now gradients of $H$, one can directly write out $I$ using the current-current correlation measure. The Liouville operator gives energy differences $E-E'$, its inverse inverses of these differences. Therefore, upon taking also the limit $\eta\to 0$, 
\begin{eqnarray*}
I
& = &
\frac{\imath}{2}\;
\int_{\RM^2}m_{j+1,j+2}(dE,dE')
\;\frac{(\mu-E)\chi(\mu-E)\chi(E'-\mu) -(\mu-E')\chi(\mu-E')\chi(E-\mu)}{(E'-E)^2}
\\
& &  \;-\,
\frac{\imath}{2}\;
\int_{\RM^2}m_{j+1,j+2}(dE,dE')
\;\frac{(\mu-E)\chi(E-\mu)\chi(\mu-E') -(\mu-E')\chi(E'-\mu)\chi(\mu-E)}{(E'-E)^2}
\;.
\end{eqnarray*}
Regrouping and evaluating terms shows that $I$ is equal to the r.h.s. of \eqref{eq-maginm} so that Proposition~\ref{prop-magTcorrect} concludes the proof.
\hfill $\Box$

\vspace{.2cm} 

\noindent {\bf Remark} Let us show that formula \eqref{eq-magT=0full} reduces to the formula found in \cite{CTVR}, more precisely the equations (44) and (45) in this work with $H$ replaced by $H-\mu$ (these authors also argue that this is necessary further below in their work). Actually, \cite{CTVR} only considers Bloch electrons and thus formulas in the quasi-momentum representation, but if one transposes their formulas (44) and (45) combined with (34) and (35) read in our notations
\begin{eqnarray*}
{M}_j(\infty,\mu)
& = &
-\,\frac{\imath}{2}
\;\Tt_\BB\bigl(\nabla_{j+1}P_\mu\,Q_\mu(\mu-H)\nabla_{j+2}P_\mu-\nabla_{j+2}P_\mu\,Q_\mu(\mu-H)\nabla_{j+2}P_\mu
\bigr)
\\
& & -\,\frac{\imath}{2}\;
\Tt_\BB\bigl((\mu-H) \nabla_{j+1}P_\mu\,Q_\mu\nabla_{j+2}P_\mu-(\mu-H)\nabla_{j+2}P_\mu\,Q_\mu\nabla_{j+1}P_\mu
\bigr)
\;,
\end{eqnarray*}
where $Q_\mu=\one-P_\mu$. Using $\nabla_{j}P_\mu\,Q_\mu=P_\mu\nabla_{j}P_\mu$ twice in the second summand and regrouping the terms then directly leads to  \eqref{eq-magT=0}.
\hfill $\diamond$

\section{Derivative of magnetization w.r.t. chemical potential}
\label{sec-derivative}

It follows immediately from Theorem~\ref{theo-magT} or formula~\eqref{eq-gsplit} that
\begin{equation}
\label{eq-magderiv}
\partial_\mu\;{M}_j({\beta},\mu)
\;=\;
\frac{\imath}{2}\,
\int m_{j+1,j+2}(dE,dE')\;
\left(
\frac{f'_{\beta,\mu}(E)+f'_{\beta,\mu}(E')}{E'-E}
\;+\;
2\;\frac{f_{\beta,\mu}(E')-f_{\beta,\mu}(E)}{(E'-E)^2}
\right)
\;.
\end{equation}
From this equation the following link of the magnetization with topological invariants can be established. 

\begin{theo}
\label{theo-magderiv}
Suppose that the chemical potential $\mu$ lies in an interval $\Delta$ with finite localization length $\ell^2(\Delta)$ and that the density of states has no atom at $\mu$.  Let $P_\mu=f_{\infty,\mu}(H)$ denote the Fermi projection and suppose that $\delta H=0$. Then, with $j$ taken cyclically,  
\begin{equation}
\label{eq-magderivT=0}
\partial_\mu\;{M}_j({\infty,\mu})
\;=\;
 \imath\;\Tt_{\BB}\bigl(P_\mu[\nabla_{j+1}P_\mu,\nabla_{j+2}P_\mu]\bigr)
\;.
\end{equation}
\end{theo}

\noindent {\bf Proof.} 
One has to take the limit of \eqref{eq-magderiv} which has to be done with some care as in the proof of Proposition~\ref{prop-magTcorrect}. In particular, the first summand in \eqref{eq-magderiv} vanishes in the limit $\beta\to\infty$ (due to dynamical localization and H\"older continuity of the density of states). Therefore
$$
\partial_\mu\;{M}_j(\infty,\mu)
\;=\;
 \imath\,
\int m_{j+1,j+2}(dE,dE')\;
\frac{\chi(\mu-E)-\chi(\mu-E')}{(E'-E)^2}
\;.
$$
Now one repeat the argument in the proof of Theorem~\ref{theo-magT=0} in order to complete the proof.
\hfill $\Box$

\vspace{.2cm}

On first sight, equation \eqref{eq-magderivT=0} is puzzeling because, even if the Fermi level $\mu$ lies in a gap of the spectrum of $H$, the r.h.s. and thus the derivative of the magnetization w.r.t. to the Fermi level can be non-vanishing. The reason is that the magnetization itself is defined as the rate of change of the pressure w.r.t. the magnetic field and the pressure in turn does vary with the chemical potential:
\begin{equation}
\label{eq-pressurederive}
\partial_\mu\;p(\beta,\mu)
\;=\;
\Tt_\BB(f_{\beta,\mu}(H))
\;.
\end{equation}
Thus also the magnetiszation varies with the chemical potential (also at zero temperature as shown in Theorem~\ref{theo-Streda} below). Moreover, these derivatives are quantized as will be discussed next. Indeed, the r.h.s. of \eqref{eq-magderivT=0} can be expressed in terms of the Chern invariants of the Fermi projection which for any $P\in H^1(\Aa_\BB,\Tt_\BB)$ are defined by
\begin{equation}
\label{eq-Chern}
\mbox{\rm Ch}_j(P)
\;=\;
2\pi\imath\;\Tt_{\BB}\bigl(P[\nabla_{j+1}P,\nabla_{j+2}P]\bigr)
\;.
\end{equation}
The following are the fundamental results on these objects. For $d=2$ and $L=1$, they are proved in \cite{BES}. We refrain from giving detailed proofs for the case $L>1$ and $d=3$ because they are completely analogous.

\begin{theo}
\label{theo-Chern1}
For any $P\in H^1(\Aa_\BB,\Tt_\BB)$, the Chern invariant $\mbox{\rm Ch}_j(P)$ is an integer number given by the Fredholm index of a certain Fredholm operator. In particular, $\mbox{\rm Ch}_j(P)$ is a homotopy invariant.
\end{theo}

\begin{theo}
\label{theo-Chern2}
Suppose that $\ell^2(\Delta)<\infty$ for some interval $\Delta$. Let $P_\mu=f_{\infty,\mu}(H)$ denote the Fermi projection. Then $\mu\in\Delta\mapsto\mbox{\rm Ch}_j(P_\mu)$ is constant.
\end{theo}

If a model has its Fermi level $\mu$ in a region of dynamical localization, it is called an insulator. Indeed, the direct conductivity vanishes in the zero-dissipation limit \cite{SBB}. If, moreover, the Chern invariant $\mbox{\rm Ch}_j(P_\mu)$ does not vanish, one speaks of a Chern insulator, which is a subclass of topological insulators. The most prominent (prototypical) example is the Haldane model \cite{Hal}. An insulator with vanishing Chern invariants is often also called a normal insulator (there are, however, also non-trivial topological insulators with vanishing Chern invariants).

\vspace{.2cm}

The Hall conductivity $\sigma_{i,j}(\beta,\mu)$, $i\neq j$, is the linear response coefficient of the dissipationless transverse current. At zero temperature and if the Fermi level lies in a region of dynamical localization, it is up to numerical constants given by the Chern invariant
$$
\sigma_{j,j+1}(\infty,\mu)
\;=\;
\mbox{\rm Ch}_{j+2}(P_\mu)
\;.
$$
For $d=2$, a proof is contained in \cite{BES}, the three-dimensional generalization will be dealt with elsewhere. It therefore follows from \eqref{eq-magderivT=0} that for $\mu$ in a region of dynamical localization
$$
\partial_\mu\;{M}_j({\infty,\mu})
\;=\;
\frac{1}{2\pi}\;
\sigma_{j+1,j+2}(\infty,\mu)
\;.
$$
The Chern invariant is related to another quantity, namely the derivative of the density of states w.r.t. the magnetic field. In the context of the quantum Hall effect, this is known as Streda's formula \cite{Str,Bel2}. 

\begin{theo}
\label{theo-Streda}
Suppose that the Fermi level $\mu$ lies in a region of dynamical localization.  Let $P_\mu=f_{\infty,\mu}(H)$ denote the Fermi projection and suppose that $\delta H=0$. Then, with $j$ taken cyclically, Streda's formula holds:
\begin{equation}
\label{eq-Streda}
\partial_{B_j}\,\Tt_\BB(P_\mu)
\;=\;
-\, \imath\;\Tt_{\BB}\bigl(P_\mu[\nabla_{j+1}P_\mu,\nabla_{j+2}P_\mu]\bigr)
\;.
\end{equation}
\end{theo}

\noindent {\bf Proof.} 
Because it is short and elucidating, let us provide a proof only in the case where $\mu$ lies in a gap of the spectrum of $H$ (the general case goes along the lines above and is again somewhat lengthy). As above, with the notation $P=P_\mu$,
$$
\partial_{B_j}\,\Tt_\BB(P)
\;=\;
\Tt_\BB(\delta_j P^2)
\;=\;
2\;\Tt_\BB(P\delta_j P)\,+\,\frac{\imath}{2}\,\Tt_{\BB}\bigl([\nabla_{j+1}P,\nabla_{j+2}P]\bigr)
\;=\;
2\;\Tt_\BB(P\delta_j P)
\;.
$$
Then \eqref{eq-diagdelta} already concludes the proof.
\hfill $\Box$

\vspace{.2cm}

As a corollary, for $\mu$ in a region of dynamical localization,
$$
\partial_\mu\;{M}_j({\infty,\mu})
\;=\;
-\;
\partial_{B_j}\,\Tt_\BB(P_\mu)
\;.
$$
This follows also if one derives the zero-temperature limit of the identity \eqref{eq-pressurederive} w.r.t. the magnetic field. Thus, one can derive Streda's formula from Theorem~\ref{theo-magderiv} and vica versa.

\section{Boundary currents in Chern insulators}
\label{sec-boundary}

In this brief section, we sketch how non-trivial Chern numbers in Chern insulators lead to edge currents, just as they do in quantum Hall systems (actually,  quantum Hall systems are special cases of Chern insulators). The results are given without detailed proofs and even without a detailed description of the set-up. In dimension $d=2$, all this is given in \cite{KRS}, while the case $d=3$ will be dealt with in a future publication. Let us consider the half-spaces $\Gamma_j=\{(n_1,\ldots,n_d)\,|\,n_j\geq 0\}$. Then $\widehat{H}_{j,\omega}$ denotes the restriction of $H_\omega=\pi_{\BB,\omega}(H)$ to $\ell^2(\Gamma_j)$ by imposing Dirichlet boundary conditions. The family $(\widehat{H}_{j,\omega})_{\omega\in\Omega}$ is still satisfies the covariance relation \eqref{eq-covariance} in the direction $i\neq j$. These operators are the representations of an operator $\widehat{H}_j$ lying in a Toeplitz extension $\Ee$ of $\Aa$ which is constructed in \cite{KRS}. This extension $\Ee$ in turn contains operators $\widehat{A}_j=(\widehat{A}_{j,\omega})_{\omega\in\Omega}$ for which the following trace is well-defined and finite:
$$
\widehat{\Tt}_j(\widehat{A}_j)
\;=\;
\EE\;\sum_{n_j\geq 0}\;
\langle n_je_j|\widehat{A}_{j,\omega}|n_je_j\rangle
\;,
$$
where $e_j$ is the $j$th unit vector so that $n_je_j$ is the $d$-dimensional vector having vanishing components except the $j$th which is equal to $n_j$. Operators $\widehat{A}_j$ for which $|\widehat{A}_j|$ has finite $\widehat{\Tt}_j$-trace are called $\widehat{\Tt}_j$-traceclass. The following result can be proved using the techniques of \cite{KRS} (details will be presented elsewhere). It shows that the boundary currents of any Chern insulator are determined by bulk properties, just as for quantum Hall systems. 

\begin{theo}
\label{theo-boundary}
Let $H$ have a gap $\Delta$ and let $F$ be a positive smooth function supported in $\Delta$ which has unit integral $\int dE\,F(E)=1$. Then $F(\widehat{H}_j)$ is $\widehat{\Tt}_j$-traceclass and for $i\neq j$ and any $\mu\in\Delta$
$$
\widehat{\Tt}_j
\bigl(F(\widehat{H}_j)\,\nabla_i(\widehat{H}_j)\bigr)
\;=\;
\mbox{\rm Ch}_{j+i}(P_\mu)
\;,
$$
where $j+i$ is calculated modulo $d$.
\end{theo}

It may be possible to extend this result to Chern insulators with a mobility gap (Fermi level in an interval of dynamical localization) if as in \cite{EGS} an adequate limit procedure for the calculation of the edge currents is chosen.

\section{Magnetic susceptibility}
\label{sec-susceptibility}

There are two alternatives for the calculation of the orbital magnetic susceptibility: either conventional linear response theory in the magnetic field, or simply derivation of the magnetization w.r.t. the magnetic field. In view of the above, it is natural to follow the second approach which has, moreover, the advantage of not needing any approximation. The magnetic susceptibility is then the tensor defined by
$$
\chi_{j,i}(\beta,\mu)
\;=\;
\partial_{B_i}\;M_{j}(\beta,\mu)
\;,
\qquad
i,j=1,\ldots,d
\;.
$$
Using \eqref{eq-Magint} it is natural to first calculate
$$
\widetilde{\chi}_{j,i}(\beta,\mu)
\;=\;
\partial_{B_i}\;\partial_{B_j}\;\Tt_\BB\bigl(f_{\beta,\mu}(H)\,(\mu-H)\bigr)
\;,
$$
because $\chi_{j,i}(\beta,\mu)$ is then obtained as average over the inverse temperature. In principle, there is no problem to calculate $\chi_{j,i}(\beta,\mu)$, because for any smooth and compactly supported function $f$ with Fourier transform $\hat{f}$:
\begin{equation}
\label{eq-f(H)Itoderiv}
\delta_i f(H)
\;=\;
\frac{1}{2\imath}\;\int dt\;\hat{f}(t)\,\int^1_0 ds\int^s_0 dr\;e^{\imath(1-s)t H}\,[\nabla_{i+1}He^{\imath rt H},
\nabla_{i+2}He^{\imath (s-r)t H}]
\;.
\end{equation}
This leads to some formula for the susceptibility only invoking the gradients $\nabla_j H$, but the algebraic complexity is considerable and it is not clear whether that formula is of any help.  On the other hand, numerous prior works (nicely reviewed in \cite{BCS}) have shown that any final expression for the susceptibility simply contains many terms so that this complexity seems unavoidable.

\appendix

\section{Adiabatic theorem for C$^*$-algebras}
\label{app-adiabatic}

The following is essentially an adaption of the super-adiabatic theorem of Nenciu \cite{Nen} for unitary evolutions on Hilbert spaces to the   setting of C$^*$-dynamical systems. All algebras and notations are as in Sections~\ref{sec-setup} and \ref{sec-Polar}, but we emphasize that the results and proofs hold for general $C^*$-dynamical systems generated by bounded $*$-derivations.  Hence let $t\in[0,T]\mapsto H(t)\in C^1_{{\rm s}}(\Aa)$ be a time-dependent Hamiltonian (the derivative $\delta$ w.r.t.\ the magnetic field plays no role here, so $\BB$ is supposed to be fixed). In the remainder of this appendix, it is always assumed that the following hypothesis holds.

\vspace{.2cm}

\noindent{\bf Gap condition:} {\it Let $\sigma_*(t)\subset \sigma(H(t))$ be a subset of the spectrum of $H(t)$ such that there exist continuous functions $f_\pm:[0,T] \to \RM$ defining  intervals $I(t)= [f_-(t),f_+(t)]$ such that $\sigma_*(t) \subset I(t)$ and
$$
g\;=\;\inf_{t\in[0,T]}\; {\rm dist}( I(t), \sigma(H(t))\setminus \sigma_*(t))
$$
is strictly positive. }

\vspace{.2cm}

\begin{lemma}
\label{lemm-pert}
Assume the gap condition and let $P(t) = \chi_{\sigma_*(t)}(H(t))$ be the spectral projection of $H(t)$ associated with $\sigma_*(t)$. Then $P(t)\in C^1_{\rm s}(\Aa)$ for all $t\in[0,T]$. If $H\in C^{N}([0,T], C^1_{\rm s}(\Aa))$, then also 
$P\in C^{N}([0,T], C^1_{\rm s}(\Aa))$. 
\end{lemma} 

\noindent {\bf Proof.} 
Let $\gamma(t)\subset  \CM$ be a closed curve in the resolvent set of $H(t)$ encircling $\sigma_*(t)$  once in the positive sense with dist$(\gamma(t), \sigma(H(t))\setminus\sigma_*(t)) \leq \frac{g}{2}$. Then 
$$
P(t) \;=\; 
\oint_{\gamma(t)} \frac{dz}{2\pi\imath}\; (z- H(t) )^{-1}\in C^1_{\rm s}(\Aa)\,,
$$
since according to Proposition~\ref{prop-diffcalculus} one has  $(z-H(t))^{-1} \in C^1_{\rm s}(\Aa)$ for all $z\not\in\RM$. The continuity of $f_{\pm}(t)$ implies that $\gamma(t+h)$ is homotopic to $\gamma(t)$ in the resolvent set of $H(t+h)$ for $|h|$ small enough and hence
$$
P(t+h)\; = \;
\oint_{\gamma(t+h)} \frac{dz}{2\pi\imath}\; (z- H(t+h) )^{-1}
\;=\; 
\oint_{\gamma(t)} \frac{dz}{2\pi\imath}\; (z- H(t+h) )^{-1}
$$
for $|h|$ small enough. Therefore
$$
\partial_t^n\; P(t) 
\;= \;
\oint_{\gamma(t)} \frac{dz}{2\pi\imath}\; \partial_t^n\;  (z- H(t) )^{-1} 
\;\in\; 
C^1_{\rm s}(\Aa)
\;,
$$
for $n\leq $, since $t\mapsto H(t)$ and thus also $t\mapsto (z- H(t))^{-1}$ is $N$ times differentiable. The latter fact follows from
$$
\partial_t\;    (z- H(t) )^{-1} 
\;=\; 
(z- H(t) )^{-1} \,\dot H(t)\, (z- H(t) )^{-1}
\;,
$$
and induction. 
\hfill $\Box$

\vspace{.2cm}

For the formulation of the following theorem, let us introduce   for any differentiable path $t\in[0,T] \mapsto A(t)\in C^1_{\rm s}(\mathcal{A})$ the abbreviation
$$
\|\!\:\!| A(t) \|\!\:\!|
 \;=\;
\| A(t) \| \,+ \,
\| \partial_t\,A(t) \| 
\,+\, 
\|\nabla A(t) \|  \;.
$$

\begin{theo}[\bf Superadiabatic projections]
 \label{theo-adiabatics}
 Let $H\in C^{N+2}([0,T], C^1_{\rm s}(\Aa))$ for some $N\in\NM_0$ and assume the gap condition. 
 Let $P(t)$ be the spectral projection of $H(t)$ associated with the gapped part of the spectrum just as in {\rm Lemma~\ref{lemm-pert}}. Then there are $\varepsilon_N>0$, $C_N<\infty$ and orthogonal projections $P^\varepsilon_N(t)\in  C^1_{\rm s}(\mathcal{A})$ such that the map $\varepsilon \mapsto P^\varepsilon_N(\cdot)\in C^2([0,T];C^1_{\rm s}(\mathcal{A}))$ is   differentiable and
 the following properties hold uniformly in $t$:
 \begin{equation}\label{ST1}
\|\!\:\!| P^\varepsilon_N(t) - P(t) \|\!\:\!|
\;\leq \;C_N \;\varepsilon \;,
 \end{equation}
\begin{equation}\label{ST2}
\|\!\:\!| 
\bigl[ \imath\,\varepsilon \partial_t - H(t), P^\varepsilon_N(t) \bigr]
\|\!\:\!|
\; \leq \;C_N \, \varepsilon^{N+1}\;.
\end{equation}
If $\partial_t^n\,H(t)|_{t=\tau}=0$ for some $\tau\in[0,T]$  and all $n\leq N$, then
$P_N^\varepsilon(\tau) = P(\tau)$.
 \end{theo}

The proof of Theorem~\ref{theo-adiabatics} is based on the following explicit construction.

\begin{proposi}
\label{prop-preparation}
There exist unique functions $P_n\in C^{N+2-n}([0,T], C^1_{\rm s}(\Aa))$, $1\leq n\leq N$, such that  
\[
 \widetilde {P}^\varepsilon_m(t) \;=\; \sum_{n=0}^m\varepsilon^n P_n(t)
\]
for $0\leq m\leq N$ with $P_0(t) \;= P(t)$  
 satisfies  
\begin{equation}\label{IA1}
 \widetilde {P}_m^\varepsilon  \widetilde {P}_m^\varepsilon  \;=\;  \widetilde {P}_m^\varepsilon \; + \;
\varepsilon^{m+1} G_{m+1}\; +\; \Oo (\varepsilon^{m+2}) 
\end{equation}
with $ G_{m+1}= \sum_{n=1}^m P_n P_{m+1-n}$ and 
\begin{equation}
\label{IA2}
 \left[ \imath\,\varepsilon \partial_t - H ,  \widetilde {P}^\varepsilon_m  \right] \;=\; \imath\,\varepsilon^{m+1}  \dot P_m \,.
\end{equation}
Here and in the following we sometimes drop the $t$-dependence in the notations to make formulas shorter.
If $\partial_t^n\,H(t)|_{t=\tau}=0$ for some $\tau\in[0,T]$  and all $n\leq N$, then
$P_n(\tau) = 0$ for all $1\leq n\leq N$.
\end{proposi}

\noindent {\bf Proof.} One constructs $P_j$ by induction in $m$.
Let $P_0 =P $, then $ \widetilde {P}_0^\varepsilon=P_0$ so that
$$
 \widetilde {P}^\varepsilon_0 \widetilde {P}^\varepsilon_0  \;=\;  \widetilde {P}^\varepsilon_0 \quad\mbox{and} \quad   \left[ \imath\,\varepsilon \partial_t - H(t),  \widetilde {P}^\varepsilon_0(t) \right] \;=\; \imath\,\varepsilon \dot P_0(t) \;=\;\Oo (\varepsilon)\,. 
$$
Now assume that $P_j$ are already constructed for $j=0,\ldots, m$ such that (\ref{IA1}) and (\ref{IA2}) hold.  One needs to  show that there is a unique smooth  $P_{m+1}(t)$ such that $\widetilde {P}_{m+1}^\varepsilon =  \widetilde {P}_m^\varepsilon + \varepsilon^{m+1} P_{m+1}$ satisfies 
\begin{equation}
\label{C1}
 \widetilde {P}_{m+1}^\varepsilon  \widetilde {P}_{m+1}^\varepsilon  \;=\;  \widetilde {P}_{m+1}^\varepsilon  + \varepsilon^{m+2} G_{m+2} + \Oo(\varepsilon^{m+3}) 
\end{equation}
and
\begin{equation}
\label{C2}
 \left[ \imath\,\varepsilon \partial_t - H(t), \widetilde{P}^\varepsilon_{m+1}(t) \right] \;=\; \imath\,\varepsilon^{m+2}  \dot P_{m+1}(t)\,.
\end{equation}
Using  (\ref{IA1}) one finds that 
\begin{eqnarray*}
\widetilde{P}_{m+1}^\varepsilon \widetilde{P}_{m+1}^\varepsilon   &=& \widetilde{P}_m^\varepsilon \widetilde{P}_m^\varepsilon   + \varepsilon^{m+1} \left(
P_0 P_{m+1} + P_{m+1} P_0 \right)+ \Oo(\varepsilon^{m+2}) \\
&=& 
 \widetilde{P}_m^\varepsilon     + \varepsilon^{m+1} \left(G_{m+1}+
P_0 P_{m+1} + P_{m+1} P_0 \right)+ \Oo(\varepsilon^{m+2}) \,.
\end{eqnarray*}
In order to obtain (\ref{C1}) up to order $\varepsilon^{m+2}$,   $P_{m+1}$ needs to be chosen such that 
\begin{equation}
\label{cond1}
G_{m+1}+
P_0 P_{m+1} + P_{m+1} P_0 \;=\; P_{m+1} \,.
\end{equation}

\vspace{.1cm}

\noindent {\bf Claim 1:} (\ref{IA1})   implies that  
$[P_0 , G_{m+1} ] = 0$. 

\vspace{.1cm}

Indeed, with $\widetilde{P}_\varepsilon - P_0 = \Oo(\varepsilon)$  it holds that
\begin{eqnarray*}
\varepsilon^{m+1} [ P_0 , G_{m+1} ] &=& \varepsilon^{m+1} [ \widetilde{P}^\varepsilon_m , G_{m+1} ] \;+\; \Oo(\varepsilon^{m+2}) \\&\stackrel{(\ref{IA1})}{=}& 
  [ \widetilde{P}^\varepsilon_m ,  (\widetilde{P}^\varepsilon_m)^2- \widetilde{P}^\varepsilon_m ]\; +\; \Oo(\varepsilon^{m+2}) \;=\; \Oo(\varepsilon^{m+2})\,.
\end{eqnarray*}
Since $[P_0, G_{m+1}]$ does not depend on $\varepsilon$, the claim follows.

\vspace{.2cm}

Claim~1 implies that  (\ref{cond1}) and thus also (\ref{C1}) holds if and only if  the diagonal blocks of $P_{m+1}$ are set equal to
\begin{equation}\label{diagdef}
P_0 P_{m+1} P_0 \;=\; -\,P_0 G_{m+1} P_0 \quad\mbox{and}\quad P_0^\perp P_{m+1} P_0^\perp \;=\;  P_0^\perp G_{m+1} P_0^\perp\,,
\end{equation}
where  $P_0^\perp= 1-P_0$.  Note that (\ref{cond1}) puts no condition on the off-diagonal blocks. To fix the latter let us evaluate the l.h.s. of (\ref{C2}) using \eqref{IA2}:
$$
 \left[ \imath\,\varepsilon \partial_t - H , \widetilde{P}^\varepsilon_{m+1}  \right] \;=\; \varepsilon^{m+1} \left(\imath \dot P_m - [H, P_{m+1}]\right) + \imath\,\varepsilon^{m+2} \dot P_{m+1}
 \;.
$$
Comparing with (\ref{C2}) let us require
\begin{equation}
\label{cond2}
\imath \dot P_m - [H, P_{m+1}] \;=\; 0\,.
\end{equation}
The following shows that the diagonal blocks  of the left hand side in~(\ref{cond2}) are automatically zero for our choice (\ref{diagdef}).

\vspace{.1cm}

\noindent {\bf Claim 2:}  (\ref{IA1}) and (\ref{IA2})   imply that 
$$
P_0\dot P_mP_0 \;=\; \imath P_0[ H,G_{m+1}] P_0\quad \mbox{and}\quad P_0^\perp \dot P_mP_0^\perp \;=\;- \imath P_0^\perp[ H,G_{m+1}] P_0^\perp\,.
$$

\vspace{.1cm}

Indeed, from
\begin{eqnarray*} 
\varepsilon^{m+1} [ H , G_{m+1} ] &=& \varepsilon^{m+1} [ H-\imath\,\varepsilon \partial_t , G_{m+1} ] + \Oo(\varepsilon^{m+2}) \\&\stackrel{(\ref{IA1})}{=}& 
  [  H-\imath\,\varepsilon \partial_t,  (\widetilde{P}^\varepsilon_m)^2- \widetilde{P}^\varepsilon_m ] + \Oo(\varepsilon^{m+2}) \\
 &\stackrel{(\ref{IA2})}{=}&  -\imath\,\varepsilon^{m+1} \left(\dot P_m P_0 + P_0 \dot P_m - \dot P_m
  \right)
  + \Oo(\varepsilon^{m+2}) 
\end{eqnarray*}
and  the fact that  $[ H , G_{m+1} ]+\imath(\dot P_m P_0 + P_0 \dot P_m - \dot P_m)$ is independent of $\varepsilon$,
it follows that 
$$
[ H , G_{m+1} ]+\imath(\dot P_m P_0 + P_0 \dot P_m - \dot P_m)\;=\;0\,.
$$
Taking the diagonal blocks of  this equality proves the claim.

\vspace{.2cm}

The off-diagonal blocks of (\ref{cond2}) vanish  if and only if 
\begin{equation}
\label{cond3}
 [ \imath \dot P_m - [H, P_{m+1}] , P_0] \;=\; 0\,.
\end{equation}
Let $z\in\CM$ be in the resolvent set of $H(t)$ and put $R_z(t)= (z-H(t))^{-1}$. Then (\ref{cond3}) is equivalent to
\begin{equation}
\label{cond4}
  R_z [\imath \dot P_m - [H, P_{m+1}] , P_0] R_z \;=\; 0\,.
\end{equation}
Now using $[R_z,P_0]=0$ one checks
$$
R_z [ [H, P_{m+1}] , P_0] R_z \;=\; R_z [ [(H-z), P_{m+1}] , P_0] R_z \;=\;[[P_0,P_{m+1}],R_z]
\;.
$$
Hence integrating (\ref{cond4}) along a curve $\gamma(t)$ encircling $\sigma_*(t)$ once in the positive sense gives
$$
\oint_{\gamma(t)} \frac{dz}{2\pi}\;R_z [ \dot P_m   , P_0] R_z 
\,=\, \oint_{\gamma(t)} \frac{dz}{2\pi\imath}  \,[R_z,[P_0,P_{m+1}]] 
 \,=\, [P_0,[P_0,P_{m+1}]] 
 \,=\,
 P_0^\perp P_{m+1} P_0 +  P_0 P_{m+1} P_0 ^\perp\,.
 $$
This fixes the off-diagonal terms in $P_{m+1}$. Summarizing, with the choice
\begin{equation}
P_{m+1} \;=\;  P_0^\perp G_{m+1} P_0^\perp \;-\;  P_0 G_{m+1} P_0  \;+\;\oint_{\gamma(t)} \frac{dz}{2\pi}
\; R_z [ \dot P_m   , P_0] R_z
\;,
\end{equation}
one has (\ref{C1}) and (\ref{C2}). 

\vspace{.1cm}

It remains to prove the final statement of Proposition~\ref{prop-preparation}.
First note that  $\dot H(\tau) = 0$ implies $\dot R_z(\tau)=0$ and  thus $\dot P_0(\tau)=0$ and   $P_1(\tau)=0$.  Again one proceeds by induction. Assume that 
  $\partial_t^nH(t)|_{t=\tau}=0$ for some $\tau\in\RM$  and all $n\leq m+1$ and that $P_n(\tau) = 0$ for all $n\leq m$. Then clearly $G_{m+1}(\tau) = 0$ by the definition of $G_{m+1}$ and 
$$
 P_{m+1}(\tau)  \;=\;   \oint_{\gamma(\tau) } \frac{dz}{2\pi}\; R_z(\tau)  [ \dot P_m(\tau)    , P_0(\tau) ] R_z(\tau)  \,.
$$
To see that $\dot P_m(\tau)=0$, note that iteration of 
\begin{eqnarray*}
 \dot P_{m }(\tau) &=&   \partial_t \oint_{\gamma(t) }
 \frac{dz}{2\pi} \,R_z [ \dot P_{m-1}   , P_0] R_z \Big|_{t=\tau} \\
 & =&  \oint_{\gamma(\tau) } 
 \frac{dz}{2\pi} \,R_z(\tau) [ \ddot P_{m-1}(\tau)   , P_0(\tau)] R_z(\tau)  \\
 &=&  \oint_{\gamma(\tau) } 
 \frac{dz}{2\pi}  \,\oint_{\gamma(\tau) } 
 \frac{d \tilde   z}{2\pi} \,R_z(\tau) [   R_{\tilde z}(\tau) [ \dddot P_{m-2}(\tau)   , P_0(\tau)] R_{\tilde z}(\tau)     , P_0(\tau)] R_z(\tau)\\[4mm]
 &=& \cdots
    \end{eqnarray*}
implies that $P_{m+1}(\tau)=0$, if $\partial_t^{m+1} P_0(\tau)=0$. This in turn follows form the assumption that  $\partial_t^nH(t)|_{t=\tau}=0$ for  all $n\leq m+1$.
\hfill $\Box$ 

\vspace{.2cm}

\noindent {\bf Proof} of Theorem~\ref{theo-adiabatics}.
It remains to turn $\widetilde{P}_N^\varepsilon(t)$ into a projection. Since by (\ref{IA1})  there is a constant $c_{N }$ such that $\| (\widetilde{P}_N^\varepsilon)^2-\widetilde{P}_N^\varepsilon\|\leq c_N\varepsilon^{N+1}$, by the spectral mapping theorem
$$
\sigma(\widetilde{P}_N^\varepsilon) 
\;\subset\;
[ - c_N\varepsilon^{N+1}, c_N\varepsilon^{N+1}] \cup [1-c_N\varepsilon^{N+1},1+c_N\varepsilon^{N+1}] \;\subset \;[-\tfrac{1}{4},\tfrac{1}{4}] \cup [\tfrac{3}{4}, \tfrac{5}{4}] 
\;,
$$
the latter for $\varepsilon\leq\varepsilon_N = (4 c_N)^{-\frac{1}{N+1}}$. Hence one can define for $\varepsilon\leq\varepsilon_N$
$$
P_N^\varepsilon  \;=\; \oint_{|z-1|\;=\;\frac{1}{2} } \hspace{-2pt}
\frac{dz}{2\pi\imath}\, (z-\widetilde{P}_N^\varepsilon )^{-1}
\;,
$$
where the integral is over the circle $|z-1|=\frac{1}{2}$ in the positive sense.  Then $\varepsilon\mapsto P^\varepsilon_N(\cdot )\in C^2([0,T];C^1_{\rm s}(\mathcal{A}))$  is  differentiable, since $\widetilde{P}_N^\varepsilon$ and thus also $(z-\widetilde{P}_N^\varepsilon )^{-1}$ is. Taking norms of
$$
P_N^\varepsilon -P \; =\; \oint_{|z-1|\;=\;\frac{1}{2} } \hspace{-2pt}
\frac{dz}{2\pi\imath}\, (z-\widetilde{P}_N^\varepsilon )^{-1}(\widetilde{P}_N^\varepsilon -P)(z-P)^{-1}
$$
and its derivative $\partial_t$ and gradient $\nabla$ implies (\ref{ST1}). Similarly, one can use
\begin{eqnarray*}
 \left[ \imath\,\varepsilon \partial_t - H ,  P^\varepsilon_N  \right] 
 &=&  \oint_{|z-1|=\frac{1}{2}} \frac{d z}{2\pi\imath}\;  \left[ \imath\,\varepsilon \partial_t  - H ,(z-\widetilde{P}_N^\varepsilon )^{-1}\right]\\
 &=&\oint_{|z-1|=\frac{1}{2}} \frac{d z}{2\pi\imath}\;   (z-\widetilde{P}_N^\varepsilon)^{-1} \left[ \imath\,\varepsilon \partial_t - H ,\widetilde{P}_N^\varepsilon\right](z-\widetilde{P}_N^\varepsilon )^{-1}\\
 &=&   \varepsilon^{N+1} \oint_{|z-1|=\frac{1}{2}} \frac{d z}{2\pi}\;   (z-\widetilde{P}_N^\varepsilon )^{-1} \dot P_N(z-\widetilde{P}_N^\varepsilon )^{-1}
\end{eqnarray*}
to  show (\ref{ST2}). The last claim follows directly from Proposition~\ref{prop-preparation}.
 \hfill $\Box$

\begin{coro} 
\label{coro-adiabatic}
Let $\eta^\varepsilon_{t,s}$ be the family of $*$-automorphisms providing the solutions of \eqref{eq-LiouvilleAdi}. Then
$$
\eta^\varepsilon_{t,0}( P^\varepsilon_N(0) ) \;=\; P_N^\varepsilon(t) \;+ \;\Delta^\varepsilon(t)
$$
with 
$$
\|\!\:\!| \Delta^\varepsilon (t)
\|\!\:\!|\
\; =\; \Oo ( \varepsilon^{N }|t|)\,.
$$
\end{coro}

\noindent {\bf Proof.}  Let $Q^\varepsilon(t) = \eta^\varepsilon_{t,0}(P_N^\varepsilon(0))$ so that $\Delta^\varepsilon(t)  = Q^\varepsilon(t)-P_N^\varepsilon(t)$. Then $\Delta^\varepsilon(0)=0$ and $\Delta^\varepsilon(t)$ satisfies the inhomogenous linear equation
$$
\partial_t\; \Delta^\varepsilon(t) \,=\,  \partial_t\, \left(  Q^\varepsilon(t)-P_N^\varepsilon(t)\right) \,=\, \frac{\imath}{\varepsilon} \;[ Q^\varepsilon(t)-P_N^\varepsilon(t),H(t)]\; + \;R(t)\,=\, \frac{\imath}{\varepsilon}\; [ \Delta^\varepsilon(t),H(t)] \;+ \;R(t)\;,
$$
with 
$$
R(t) \;=\; \frac{\imath}{\varepsilon} \left[ \imath\,\varepsilon\, \partial_t- H(t), P^\varepsilon_N(t) \right]
\,.
$$
Hence by the variation of constants formula  
$$
\Delta^\varepsilon(t) \;=\; \eta^\varepsilon_{t,0}\left( \int_0^td s\;  \eta^{\varepsilon }_{0,s} \left(R(s)\right)\right)\,.
$$
As $\|\!\:\!|R(t)\|\!\:\!|\leq C_N\varepsilon^{N}$ by Theorem~\ref{theo-adiabatics}, taking norms then gives $\|\!\:\!|\Delta^\varepsilon(t)\|\!\:\!|\leq |t|  C_N \varepsilon^N$.
\hfill $\Box$

\end{document}